\documentclass[
reprint,
superscriptaddress,
aps,
pre,
]{revtex4-1}

\usepackage{graphicx}
\usepackage{dcolumn}
\usepackage{bm}
\usepackage[utf8]{inputenc}
\usepackage{amsmath, amsthm, amssymb, amsfonts}
\usepackage{accents}
\usepackage[usenames,dvipsnames]{xcolor}
\usepackage{tikz}
\usepackage{array}
\usepackage{dsfont}
\usepackage{mathtools}
\usepackage{soul,xcolor}
\usepackage[colorlinks=true,linkcolor=Blue,citecolor=Blue,urlcolor=Blue]{hyperref}
\usepackage[caption=false]{subfig}
\usepackage[capitalise]{cleveref}

\pgfdeclarelayer{bg}    % declare background layer
\pgfsetlayers{bg,main}  % set the order of the layers (main is the standard layer)

\usetikzlibrary{shapes.geometric}
\usetikzlibrary{shapes.misc}
\usetikzlibrary{arrows.meta}

\newcolumntype{M}[1]{>{\centering\arraybackslash}m{#1}}

\newcommand{\YZ}{\color{black}}

\setstcolor{blue}

\begin{document}

\newcommand{\mytitle}{Topological Control of Synchronization Patterns: Trading Symmetry for Stability}
\title{\mytitle}

\author{Joseph D. Hart}
\affiliation{Institute for Research in Electronics and Applied Physics, University of Maryland, College Park, Maryland 20742, USA}
\affiliation{Department of Physics, University of Maryland, College Park, Maryland 20742, USA}
\author{Yuanzhao Zhang}
\affiliation{Department of Physics and Astronomy, Northwestern University, Evanston, Illinois 60208, USA}
\author{Rajarshi Roy}
\affiliation{Institute for Research in Electronics and Applied Physics, University of Maryland, College Park, Maryland 20742, USA}
\affiliation{Department of Physics, University of Maryland, College Park, Maryland 20742, USA}
\affiliation{Institute for Physical Science and Technology, University of Maryland, College Park, Maryland 20742, USA}
\author{Adilson E. Motter}
\affiliation{Department of Physics and Astronomy, Northwestern University, Evanston, Illinois 60208, USA}
\affiliation{Northwestern Institute on Complex Systems, Northwestern University, Evanston, Illinois 60208, USA}

%\date{10 a.m. \today}

\begin{abstract}
Symmetries are ubiquitous in network systems and have profound impacts on the observable dynamics.
At the most fundamental level, many synchronization patterns are induced by underlying network symmetry, and a high degree of symmetry is believed to enhance the stability of identical synchronization.
Yet, here we show that the synchronizability of almost any symmetry cluster in a network of identical nodes can be enhanced precisely by breaking its structural symmetry.
This counterintuitive effect holds for generic node dynamics and arbitrary network structure and is, moreover, robust against noise and imperfections typical of real systems, which we demonstrate by implementing a state-of-the-art optoelectronic experiment. 
These results lead to new possibilities for the topological control of synchronization patterns, which we substantiate by presenting an algorithm that optimizes the structure of individual clusters under various constraints. 
\vspace{3mm}

\noindent DOI: \href{https://doi.org/10.1103/PhysRevLett.122.058301}{10.1103/PhysRevLett.122.058301} 
%\hspace{10mm} Subject Areas: Complex Systems, Nonlinear Dynamics
\end{abstract}

\maketitle

Symmetry and synchronization are interrelated concepts in network systems. 
Synchronization, being a symmetric state among oscillators, has its existence and stability influenced by the symmetry of the network \cite{stewart2003symmetry,nicosia2013remote,aguiar2011dynamics}. 
For example, recent research has shown that network symmetry can be systematically explored to identify stable synchronization patterns in complex networks \cite{pecora2014cluster}.
Different work has shown that structural homogeneity (and hence a higher degree of network symmetry) usually enhances synchronization stability \cite{donetti2005entangled,denker2004breaking,nishikawa2003heterogeneity}. 
Any given network of identical oscillators can always be partitioned into so-called symmetry clusters \cite{macarthur2008symmetry}, characterized as clusters of oscillators that are identically coupled, both within the cluster and to the rest of the network, making them natural candidates for cluster synchronization \cite{pecora2014cluster,sorrentino2016complete}.
Cluster synchronization has been investigated in numerous experimental systems, including networks of optoelectronic oscillators \cite{pecora2014cluster,sorrentino2016complete,williams2013experimental}, semiconductor lasers \cite{nixon2012controlling,argyris2016experimental}, Boolean systems \cite{rosin2013control}, neurons \cite{vardi2012synchronization}, slime molds \cite{takamatsu2001spatiotemporal}, and chemical oscillators \cite{totz2015phase}.
Many of these experiments explicitly investigated the beneficial impact of network symmetries on cluster formation \cite{totz2015phase,takamatsu2001spatiotemporal,pecora2014cluster,sorrentino2016complete,hart2016experimental}. 
Taken together, previous results support the expectation that oscillators that are indistinguishable on structural grounds are also more likely to exhibit indistinguishable (synchronous) dynamics.

In this Letter, we investigate the relation between symmetry and synchronization in the general context of cluster synchronization (including global synchronization). 
We show that, in order to induce stable synchronization, one often has to break the underlying structural symmetry.
This counterintuitive result holds for the general class of networks of diffusively coupled identical oscillators with a bounded and connected stability region, and it follows rigorously from our demonstration that almost all 
%networks (or subnetworks) 
clusters exhibiting optimal synchronizability are necessarily %heterogeneous and hence 
asymmetric.
In particular, the synchronizability of almost any 
%nonintertwined 
symmetry cluster can be enhanced precisely by breaking the internal structural symmetry of the cluster. 
%This is demonstrated for arbitrary networks and also for the special case in which the entire network consists of a single symmetry cluster.
These findings add an important new dimension to the recent discovery of {\it parametric} asymmetry-induced synchronization \cite{PhysRevLett.117.114101,zhang2017asymmetry,zhang2017nonlinearity}, a scenario in which the synchronization of identically coupled identical oscillators is enhanced by assigning nonidentical parameters to the oscillators. 
Here, we show that synchronization of identically coupled identical oscillators is enhanced by changing the connection patterns of the oscillators to be nonidentical. 
We refer to this effect as {\it structural} asymmetry-induced synchronization (AISync).
We confirm that this behavior is robust against noise and can be found in real systems by providing the first experimental demonstration of structural AISync using networks of coupled optoelectronic oscillators. 
In excellent agreement with theory, the experiments show unequivocally that 
%networks can be optimized for both global and cluster synchronization
both intertwined and nonintertwined clusters can be optimized by reducing structural symmetry.

We consider a %connected 
network of $n$ diffusively coupled identical oscillators,
\begin{equation}
    \dot{\bm{x}_i} = \bm{f}(\bm{x}_i) - \sigma \sum_{j=1}^n L_{ij} \bm{h}(\bm{x}_j), 
\end{equation}
where $\bm{x}_i$ is the state of the $i$th oscillator, $\bm{f}$ is the vector field governing the uncoupled dynamics of each oscillator, 
$\bm{L} = \{L_{ij}\}$ is the Laplacian matrix describing the structure of an arbitrary  unweighed %undirected 
network, $\bm{h}$ is the interaction function, and $\sigma>0$ is the 
coupling strength.
We are interested in the dynamics inside a symmetry cluster.
To facilitate presentation, we first assume that the cluster is nonintertwined \cite{pecora2014cluster,cho2017stable}; that is, it can synchronize independent of whether other clusters synchronize or not. 
%Nonintertwined clusters are desirable in practice as they support more synchronization patterns thanks to the flexibility offered by the independence between clusters.
%The important case of intertwined clusters is discussed at the end of the this Letter.
The general case of intertwined clusters---in which desynchronization in one cluster can lead to loss of synchrony in another cluster---requires considering the intertwined clusters concurrently, and this important case is addressed after our analysis of nonintertwined clusters.

Numbering the oscillators in that cluster from $1$ to $m$, we obtain the dynamical equation for the cluster:
\begin{equation}
    \begin{split}
    \dot{\bm{x}_i} 
    & = \bm{f}(\bm{x}_i) - \sigma \sum_{j=1}^m L_{ij} \bm{h}(\bm{x}_j) + \sigma \sum_{j=m+1}^n A_{ij} \bm{h}(\bm{x}_j) \\
    & = \bm{f}(\bm{x}_i) - \sigma \sum_{j=1}^m L_{ij} \bm{h}(\bm{x}_j) + \sigma \bm{I}\big(\{\bm{x}_j\}_{j>m}\big),
    \end{split}
    \label{eq:1}
\end{equation}
where $L_{ij} = \delta_{ij}\mu_i - A_{ij}$, $\bm{A} = \{A_{ij}\}$ is the adjacency matrix of the network, $\mu_i$ is the indegree of node $i$, and the equation holds for $1 \leq i \leq m$. 
Here, we denote the input term from the rest of the network $\sum_{j=m+1}^n A_{ij} \bm{h}(\bm{x}_j)$ by $\bm{I}\big(\{\bm{x}_j\}_{j>m}\big)$ to emphasize that this term is independent of $i$ and hence equal  for all oscillators $1,\dots, m$. 
This term is zero only when the cluster receives no connection from the rest of the network, such as the important case in which the entire network consists of a single symmetry cluster (i.e., $m=n$).

For $m<n$, if we regard the cluster subnetwork consisting of oscillators $1,\dots, m$ as a separate network (by ignoring its connections with other clusters), then its $m \times m$ Laplacian matrix $\widetilde{\bm{L}}$ is closely related to the corresponding block of the $n \times n$ Laplacian matrix $\bm{L}$ of the full network:
\begin{equation}
    L_{ij} = 
    \begin{cases}
    \widetilde{L}_{ij}, & \quad 1 \leq i \neq j \leq m, \\
    \widetilde{L}_{ij} + \widetilde{\mu}, & \quad 1 \leq i = j \leq m, \\
    \end{cases}
\end{equation}
where $\widetilde{\mu} \ge 0$ is the number of connections each oscillator in the cluster receives from the rest of the network.
It is then clear that there are two differences in the dynamical equation when the cluster subnetwork is part of a larger network [i.e., as a symmetry cluster, described by \cref{eq:1}] rather than as an isolated network.
First, the Laplacian matrix $\widetilde{\bm{L}}$ in the dynamical equation is replaced by $\widehat{\bm{L}} = \{L_{ij}\}_{1\leq i,j \leq m} = \widetilde{\bm{L}} + \widetilde{\mu} \mathds{1}_m$; that is, the diagonal entries are uniformly increased by $\widetilde{\mu}$.  
Second, each oscillator now receives a common input $\sigma \bm{I}\big(\{x_j\}_{j>m}\big)$ produced by its coupling with other clusters, which generally alters the synchronization trajectory $\bm{s}_I\equiv \bm{x}_1=\cdots=\bm{x}_m$, causing it to be typically different from the ones generated by the uncoupled dynamics $\dot{\bm{s}} = \bm{f}(\bm{s})$. 
This has to be accounted for when calculating the maximum Lyapunov exponent transverse to the cluster synchronization manifold to determine the stability of the cluster synchronous state.

Despite these differences, a diagonalization procedure similar to the one used in the master stability function approach \cite{pecora1998master} can still be applied to the variational equation in order to assess the cluster's synchronization stability.
The variational equation describing the evolution of the deviation away from $\bm{s}_I$ inside the cluster can be written as
\begin{equation}
  \delta\dot{\bm{X}} = \left[ \mathds{1}_{m} \otimes J\bm{f}(\bm{s}_I) - \sigma \widehat{\bm{L}} \otimes J\bm{h}(\bm{s}_I) \right] \delta\bm{X},
  \label{eq:2}
\end{equation}
where
$\delta\bm{X} = (\delta\bm{x}_1^\intercal,\cdots,\delta\bm{x}_m^\intercal)^\intercal = (\bm{x}_1^\intercal - \bm{s}_I^\intercal,\cdots,\bm{x}_m^\intercal - \bm{s}_I^\intercal)^\intercal$ and $ \otimes$ denotes the Kronecker product.
The rest of the network does not enter the equation explicitly, other than through its influence on the coupling matrix $\widehat{\bm{L}}$ and the synchronization trajectory $\bm{s}_I$. 
If $\widehat{\bm{L}}$ is diagonalizable (as for any undirected network), the decoupling of \cref{eq:2} results in $m$ independent $d$-dimensional equations corresponding to individual perturbation modes:
\begin{equation}
    \dot{\bm{\eta}}_i = \big[ J\bm{f}(\bm{s}_I) - \sigma \widehat{v}_i J\bm{h}(\bm{s}_I) \big] \bm{\eta}_i,
    \label{eq:3}
\end{equation}
where $d$ is the dimension of node dynamics, $J$  is the Jacobian operator, $\bm{\eta} = (\bm{\eta}_1^\intercal,\cdots,\bm{\eta}_m^\intercal)^\intercal$ is $\delta\bm{X}$ expressed in the new coordinates that diagonalize $\widehat{\bm{L}}$, and $\widehat{v}_i = \widetilde{v}_i + \widetilde{\mu}$ 
are the eigenvalues of $\widehat{\bm{L}}$ in ascending order of their real parts [with $\left\{\widetilde{v}_i\right\}= \mbox{eig}(\widetilde{\bm{L}})$]. 
If $\widehat{\bm{L}}$ is not diagonalizable \cite{nishikawa2006maximum}, the analysis can be carried out by using the Jordan canonical form of this matrix to replace diagonalization by block diagonalization, as explicitly shown in the Supplemental Material \cite{SM}. 
In both cases the cluster synchronous state is stable if  $\Lambda (\sigma \widehat{v}_i )<0$ for $i=2,\dots, m$, where $\Lambda$ is the largest Lyapunov exponent of \cref{eq:3} and  $\widehat{v}_2, \dots, \widehat{v}_m$ represent the transverse modes;
the maximum transverse Lyapunov exponent (MTLE) determining the stability of the synchronous state is $\max_{2\le i\le m} \Lambda (\sigma \widehat{v}_i )$.
Moreover, for the large class of oscillator networks for which
%there is a single bounded stability region
the stability region is bounded and connected
\cite{barahona2002synchronization,li2010consensus,flunkert2010synchronizing,huang2009generic}, as assumed here {\YZ and verified for all models we consider} \cite{comment1}, the synchronizability of a cluster can be quantified in terms of the eigenratio $R = \text{Re}(\widetilde{v}_m)/\text{Re}(\widetilde{v}_2)$:  the smaller this ratio, in general, the larger the range of $\sigma$ over which the cluster synchronous state can be stable.  
The cluster subnetwork is most synchronizable when $\widetilde{v}_2 = \cdots = \widetilde{v}_m$, which also implies that all eigenvalues are real and in fact integers if the network is unweighted \cite{nishikawa2010network}, as considered here. 
It is important to notice that the optimality of the cluster subnetwork is conserved in the sense that if $\widetilde{v}_2 = \cdots = \widetilde{v}_m$ for the isolated cluster, then $\widehat{v}_2 = \cdots = \widehat{v}_m$ will hold for the cluster as part of a larger network.
%This analysis also holds for discrete-time systems, such as the ones we consider below.
Since the analysis above does not invoke the continuity of the equations anywhere, it holds for discrete-time systems as well. In this case one can simply replace $\delta \dot{\bm{X}}$ and $\delta \bm{X}$ in \cref{eq:2} by $\delta \bm{X}(t+1)$ and $\delta \bm{X}(t)$, respectively.

Now we can compare symmetry clusters with optimal clusters and show rigorously that almost all optimally synchronizable clusters are asymmetric.
Without loss of generality, we consider an  unweighted cluster in isolation and assume it has $m$ nodes and $\ell$ directed links internal to the cluster.
In a symmetry cluster, because the nodes are structurally identical, the in- and outdegrees of all nodes must be equal.
Thus, $\ell$ must be divisible by $m$ if the cluster is symmetric. 
In an optimal cluster, because $\widetilde{v}_2 = \cdots = \widetilde{v}_m\equiv \widetilde{v}$ and thus tr$(\widetilde{\bm{L}}) = (m-1)\widetilde{v}$, it follows that $\widetilde{v}= \ell/(m-1)$. 
The fact that $\widetilde{v}$ is an integer implies that $\ell$ must be divisible by $m-1$ if the cluster is optimal. 
Since $\ell \le m(m-1)$, the two divisibility conditions can be satisfied simultaneously if and only if $\ell = m(m-1)$ (i.e., when the isolated cluster is a complete graph). 
But there are numerous optimal clusters for $\ell < m(m-1)$ \cite{nishikawa2006maximum,nishikawa2010network}. 
Therefore, for any given number $m$ of nodes, all optimal clusters other than the complete graph are necessarily asymmetric, meaning that (with the exception of the complete graph) the synchronization stability of any symmetry cluster can be improved by breaking its structural symmetry \cite{comment}.
This general conclusion forms the basis of structural AISync and holds, in particular, when an entire network consists of a single symmetry cluster.

\begin{table}[t]
\begin{center}
\begin{tabular}{ M{1.55cm} | M{1.25cm} | M{1.25cm} | M{1.25cm} | M{1.25cm} | M{1.25cm} }
Symmetry \newline clusters \vspace{1mm}
    &
    \resizebox{.07\textwidth}{!}{
    \begin{tikzpicture}[
      vertex/.style={draw,circle,very thick,minimum size=.6cm},
      arc/.style={draw,very thick,
      bend left=0}]
      \node[vertex,fill=magenta!60] (p1) at (.55,1) {};
      \node[vertex,fill=magenta!60] (p3) at (-1.2,0) {};
      \node[vertex,fill=magenta!60] (p5) at (.55,-1) {};
      \node[vertex,fill=magenta!60] (p2) at (-.55,1) {};
      \node[vertex,fill=magenta!60] (p4) at (-.55,-1) {};
      \node[vertex,fill=magenta!60] (p6) at (1.2,0) {};   
      \foreach [count=\r] \row in 
      {{0,1,0,0,0,1},
       {1,0,1,0,0,0},
       {0,1,0,1,0,0},
       {0,0,1,0,1,0},
       {0,0,0,1,0,1},
       {1,0,0,0,1,0}}
      {
          \foreach [count=\c] \cell in \row
          {
              \ifnum\cell=1
                  \draw[arc] (p\r) edge (p\c);
              \fi
          }
      }
      \end{tikzpicture}
      }
      &
      \resizebox{.07\textwidth}{!}{
    \begin{tikzpicture}[
      vertex/.style={draw,circle,very thick,minimum size=.6cm},
      arc/.style={draw,very thick,
      bend left=0}]
      \node[vertex,fill=magenta!60] (p1) at (.55,1) {};
      \node[vertex,fill=magenta!60] (p3) at (-1.2,0) {};
      \node[vertex,fill=magenta!60] (p5) at (.55,-1) {};
      \node[vertex,fill=magenta!60] (p2) at (-.55,1) {};
      \node[vertex,fill=magenta!60] (p4) at (-.55,-1) {};
      \node[vertex,fill=magenta!60] (p6) at (1.2,0) {};   
      \foreach [count=\r] \row in 
      {{0,1,1,0,1,0},
       {1,0,0,1,0,1},
       {1,0,0,1,1,0},
       {0,1,1,0,0,1},
       {1,0,1,0,0,1},
       {0,1,0,1,1,0}}
      {
          \foreach [count=\c] \cell in \row
          {
              \ifnum\cell=1
                  \draw[arc] (p\r) edge (p\c);
              \fi
          }
      }
      \end{tikzpicture}
      }
      &
      \resizebox{.07\textwidth}{!}{
    \begin{tikzpicture}[
      vertex/.style={draw,circle,very thick,minimum size=.6cm},
      arc/.style={draw,very thick,
      bend left=0}]
      \node[vertex,fill=magenta!60] (p1) at (.55,1) {};
      \node[vertex,fill=magenta!60] (p3) at (-1.2,0) {};
      \node[vertex,fill=magenta!60] (p5) at (.55,-1) {};
      \node[vertex,fill=magenta!60] (p2) at (-.55,1) {};
      \node[vertex,fill=magenta!60] (p4) at (-.55,-1) {};
      \node[vertex,fill=magenta!60] (p6) at (1.2,0) {};   
      \foreach [count=\r] \row in 
      {{0,1,0,1,0,1},
       {1,0,1,0,1,0},
       {0,1,0,1,0,1},
       {1,0,1,0,1,0},
       {0,1,0,1,0,1},
       {1,0,1,0,1,0}}
      {
          \foreach [count=\c] \cell in \row
          {
              \ifnum\cell=1
                  \draw[arc] (p\r) edge (p\c);
              \fi
          }
      }
      \end{tikzpicture}
      }
      &
      \resizebox{.07\textwidth}{!}{
    \begin{tikzpicture}[
      vertex/.style={draw,circle,very thick,minimum size=.6cm},
      arc/.style={draw,very thick,
      bend left=0}]
      \node[vertex,fill=magenta!60] (p1) at (.55,1) {};
      \node[vertex,fill=magenta!60] (p3) at (-1.2,0) {};
      \node[vertex,fill=magenta!60] (p5) at (.55,-1) {};
      \node[vertex,fill=magenta!60] (p2) at (-.55,1) {};
      \node[vertex,fill=magenta!60] (p4) at (-.55,-1) {};
      \node[vertex,fill=magenta!60] (p6) at (1.2,0) {};   
      \foreach [count=\r] \row in 
      {{0,0,1,1,1,1},
       {0,0,1,1,1,1},
       {1,1,0,0,1,1},
       {1,1,0,0,1,1},
       {1,1,1,1,0,0},
       {1,1,1,1,0,0}}
      {
          \foreach [count=\c] \cell in \row
          {
              \ifnum\cell=1
                  \draw[arc] (p\r) edge (p\c);
              \fi
          }
      }
      \end{tikzpicture}
      }
      &
      \resizebox{.07\textwidth}{!}{
    \begin{tikzpicture}[
      vertex/.style={draw,circle,very thick,minimum size=.6cm},
      arc/.style={draw,very thick,
      bend left=0}]
      \node[vertex,fill=magenta!60] (p1) at (.55,1) {};
      \node[vertex,fill=magenta!60] (p3) at (-1.2,0) {};
      \node[vertex,fill=magenta!60] (p5) at (.55,-1) {};
      \node[vertex,fill=magenta!60] (p2) at (-.55,1) {};
      \node[vertex,fill=magenta!60] (p4) at (-.55,-1) {};
      \node[vertex,fill=magenta!60] (p6) at (1.2,0) {};   
      \foreach [count=\r] \row in 
      {{0,1,1,1,1,1},
       {1,0,1,1,1,1},
       {1,1,0,1,1,1},
       {1,1,1,0,1,1},
       {1,1,1,1,0,1},
       {1,1,1,1,1,0}}
      {
          \foreach [count=\c] \cell in \row
          {
              \ifnum\cell=1
                  \draw[arc] (p\r) edge (p\c);
              \fi
          }
      }
      \end{tikzpicture}
      }\\ \hline
      %nontrivial \newline eigenvalues & 
      %1,1,3,3,4 & 
      %2,3,3,5,5 &
      %3,3,3,3,6 &
      %4,4,4,6,6 &
      %6,6,6,6,6 \\ \hline
      Eigenratio & 
      % beware the notation conflict
      4 & 
      2.5 &
      2 &
      1.5 &
      1 \\ \hline
    Optimal \newline clusters \vspace{1mm}
    &
    \resizebox{.07\textwidth}{!}{
    \begin{tikzpicture}[
      vertex/.style={draw,circle,very thick,minimum size=.6cm},
      arc/.style={draw,very thick,
      -{Latex[length=3mm, width=2mm]},
      bend left=0}]
      \node[] (p0) at (0,1.4) {};
      \node[vertex,fill=magenta!60] (p1) at (.55,1) {};
      \node[vertex,fill=magenta!60] (p3) at (-1.2,0) {};
      \node[vertex,fill=magenta!60] (p5) at (.55,-1) {};
      \node[vertex,fill=magenta!60] (p2) at (-.55,1) {};
      \node[vertex,fill=magenta!60] (p4) at (-.55,-1) {};
      \node[vertex,fill=magenta!60] (p6) at (1.2,0) {};   
      \foreach [count=\r] \row in 
      {{0,0,0,0,0,0},
       {1,0,0,0,0,0},
       {0,1,0,0,0,0},
       {0,0,1,0,0,0},
       {0,0,0,1,0,0},
       {0,0,0,0,1,0}}
      {
          \foreach [count=\c] \cell in \row
          {
              \ifnum\cell=1
                  \draw[arc] (p\c) edge (p\r);
              \fi
          }
      }
      \end{tikzpicture}
      }
      &
      \resizebox{.07\textwidth}{!}{
    \begin{tikzpicture}[
      vertex/.style={draw,circle,very thick,minimum size=.6cm},
      arc/.style={draw,very thick,
      -{Latex[length=3mm, width=2mm]}}]
      \node[] (p0) at (0,1.4) {};
      \node[vertex,fill=magenta!60] (p1) at (.55,1) {};
      \node[vertex,fill=magenta!60] (p3) at (-1.2,0) {};
      \node[vertex,fill=magenta!60] (p5) at (.55,-1) {};
      \node[vertex,fill=magenta!60] (p2) at (-.55,1) {};
      \node[vertex,fill=magenta!60] (p4) at (-.55,-1) {};
      \node[vertex,fill=magenta!60] (p6) at (1.2,0) {};   
      \foreach [count=\r] \row in 
      {{0,2,1,0,0,0},
       {2,0,0,0,0,3},
       {0,0,0,1,1,0},
       {0,3,0,0,0,0},
       {1,0,0,0,0,1},
       {0,0,0,3,0,0}}
      {
          \foreach [count=\c] \cell in \row
          {
              \ifnum\cell=1
                  \draw[arc,bend right=10] (p\c) edge (p\r);
              \fi
              \ifnum\cell=2
                  \draw[very thick] (p\c) edge (p\r);
              \fi
              \ifnum\cell=3
                  \draw[arc,bend left=10] (p\c) edge (p\r);
              \fi
          }
      }
      \end{tikzpicture}
      }
      &
      \resizebox{.07\textwidth}{!}{
    \begin{tikzpicture}[
      vertex/.style={draw,circle,very thick,minimum size=.6cm},
      arc/.style={draw,very thick,
      -{Latex[length=3mm, width=2mm]},
      bend left=0}]
      \node[] (p0) at (0,1.4) {};
      \node[vertex,fill=magenta!60] (p1) at (.55,1) {};
      \node[vertex,fill=magenta!60] (p3) at (-1.2,0) {};
      \node[vertex,fill=magenta!60] (p5) at (.55,-1) {};
      \node[vertex,fill=magenta!60] (p2) at (-.55,1) {};
      \node[vertex,fill=magenta!60] (p4) at (-.55,-1) {};
      \node[vertex,fill=magenta!60] (p6) at (1.2,0) {};   
      \foreach [count=\r] \row in 
      {{0,0,0,0,0,0},
       {1,0,0,0,0,0},
       {0,1,0,0,0,0},
       {0,0,1,0,0,0},
       {0,0,0,1,0,0},
       {0,0,0,0,1,0}}
      {
          \foreach [count=\c] \cell in \row
          {
              \ifnum\cell=1
                  \draw[arc] (p\c) edge (p\r);
              \fi
          }
      }
      \end{tikzpicture}
      }
      &
      \resizebox{.07\textwidth}{!}{
    \begin{tikzpicture}[
      vertex/.style={draw,circle,very thick,minimum size=.6cm},
      arc/.style={draw,very thick,
      -{Latex[length=3mm, width=2mm]},
      bend left=0}]
      \node[] (p0) at (0,1.4) {};
      \node[vertex,fill=magenta!60] (p1) at (.55,1) {};
      \node[vertex,fill=magenta!60] (p3) at (-1.2,0) {};
      \node[vertex,fill=magenta!60] (p5) at (.55,-1) {};
      \node[vertex,fill=magenta!60] (p2) at (-.55,1) {};
      \node[vertex,fill=magenta!60] (p4) at (-.55,-1) {};
      \node[vertex,fill=magenta!60] (p6) at (1.2,0) {};   
      \foreach [count=\r] \row in 
      {{0,0,3,2,0,0},
       {0,0,2,0,0,4},
       {0,2,0,0,3,2},
       {2,1,0,0,1,0},
       {1,1,0,0,0,0},
       {1,0,2,4,0,0}}
      {
          \foreach [count=\c] \cell in \row
          {
              \ifnum\cell=1
                  \draw[arc] (p\c) edge (p\r);
              \fi
              \ifnum\cell=2
                  \draw[very thick] (p\c) edge (p\r);
              \fi
              \ifnum\cell=3
                  \draw[arc, bend right=10] (p\c) edge (p\r);
              \fi
              \ifnum\cell=4
                  \draw[arc, bend left=10] (p\c) edge (p\r);
              \fi
          }
      }
      \end{tikzpicture}
      }
      &
      \resizebox{.07\textwidth}{!}{
    \begin{tikzpicture}[
      vertex/.style={draw,circle,very thick,minimum size=.6cm},
      arc/.style={draw,very thick,
      bend left=0}]
      \node[] (p0) at (0,1.4) {};
      \node[vertex,fill=magenta!60] (p1) at (.55,1) {};
      \node[vertex,fill=magenta!60] (p3) at (-1.2,0) {};
      \node[vertex,fill=magenta!60] (p5) at (.55,-1) {};
      \node[vertex,fill=magenta!60] (p2) at (-.55,1) {};
      \node[vertex,fill=magenta!60] (p4) at (-.55,-1) {};
      \node[vertex,fill=magenta!60] (p6) at (1.2,0) {};   
      \foreach [count=\r] \row in 
      {{0,1,1,1,1,1},
       {1,0,1,1,1,1},
       {1,1,0,1,1,1},
       {1,1,1,0,1,1},
       {1,1,1,1,0,1},
       {1,1,1,1,1,0}}
      {
          \foreach [count=\c] \cell in \row
          {
              \ifnum\cell=1
                  \draw[arc] (p\r) edge (p\c);
              \fi
          }
      }
      \end{tikzpicture}
      }\\ \hline
      %nontrivial \newline eigenvalues & 
      %1,1,1,1,1 & 
      %2,2,2,2,2 &
      %1,1,1,1,1 &
      %3,3,3,3,3 &
      %6,6,6,6,6 \\ 
      Eigenratio & 
      % beware the notation conflict
      1 & 
      1 &
      1 &
      1 &
      1 \\
    \end{tabular}
\vspace{-1mm}
\caption{
Connected symmetry clusters of 6 nodes and optimal clusters embedded within them.
Some symmetry clusters have more than one embedded optimal network, in which case we show one that can be obtained through a minimal number of link deletions.
}
\label{tbl:tbl1}
\end{center}
\end{table}

When viewed as isolated subnetworks, symmetry clusters are equivalent to the vertex-transitive digraphs in algebraic graph theory, defined as directed graphs in which every pair of nodes 
is equivalent under some node permutation 
\cite{biggs1993algebraic,mckay2014practical}. 
Thus, in order to improve the synchronizability of any nonintertwined symmetry cluster from an arbitrary network, we only need to optimize the corresponding vertex-transitive digraph by manipulating
its (internal) links. In particular, this can always be done by removing links inside the symmetry cluster \cite{nishikawa2006synchronization,nishikawa2010network}, despite the fact that sparser networks are usually harder to synchronize.  
For concreteness, we focus on clusters that are initially undirected and consider the selective removal of individual directional links. 
As an example, we show in \cref{tbl:tbl1} all  connected undirected symmetry clusters of 6 nodes and their embedded optimal networks. 
Apart from the complete graph, which is already optimal to begin with, the synchronizability of the other symmetry clusters as measured by the eigenratio $R$ is significantly improved in all cases.

Because in practice it can be costly or unnecessary to fully optimize a symmetry cluster, it is natural to ask whether its synchronizability can be significantly improved by just 
modifying
%\st{modifying} {\YZ rewiring} 
a few links.
%{\YZ For this purpose, link rewiring can be more effective than link removal.}
We developed an efficient algorithm for this purpose and summarize the statistical results based on all connected undirected symmetry clusters of sizes between $m=8$ and $17$ in the Supplemental Material \cite{SM}.
%Given that homogeneous networks have been shown to consistently promote synchronization, one might expect symmetry clusters to be ``locally optimal'' and hard to optimize in practice.
%Surprisingly, on average
On average, only about $14\%$ of the links need to be rewired to reduce $R-1$ by half and thus significantly improve synchronizability of symmetry clusters. 
This illustrates the potential of structural AISync as a mechanism for the topological control of synchronization stability.
Our simulated annealing code to improve cluster synchronizability is available in Ref.~\cite{SA}. 
This algorithm can also be used to demonstrate structural AISync in global synchronization, as shown in the Supplemental Material~\cite{SM}.

Having established a theoretical foundation for our main finding, we now turn to our experimental results. 
The experiments are performed using networks of identical optoelectronic oscillators whose nonlinear component is a Mach-Zehnder intensity modulator.  
The system can be modeled as
\begin{equation}
x_i(t+1)=\beta I[x_i(t)] - \sigma\sum_{j=1}^n L_{ij}I[x_j(t)] \;\; \text{mod} \;\; 2\pi,
\label{eq:exp}
\end{equation}
where $t$ is now a discrete time, $\beta$ is the feedback strength,
$I(x_i)=\sin^2(x_i+\delta)$ is the normalized intensity output of the modulator,  
$x_i$ is the normalized voltage applied to the modulator, and $\delta$ is the operating point (set to $\pi/4$ in our experiments).
Each oscillator consists of a clocked optoelectronic feedback loop. 
Light from a $780$ nm continuous-wave laser passes through the modulator, which provides the nonlinearity. 
The light intensity is converted into an electrical signal by a photoreceiver and measured by a field-programmable gate array (FPGA) via an analog-to-digital converter (ADC). 
The FPGA is clocked at 10 kHz, resulting in the discrete-time map dynamics of the oscillators.
The FPGA controls a digital-to-analog converter (DAC) that drives the modulator with a voltage $x_i(t+1)=\beta I[x_i(t)]$, closing the feedback loop.
The oscillators are coupled together electronically on the FPGA according to the desired Laplacian matrix.
Specifically, the experimental system uses time multiplexing and time delays to realize a network of coupled oscillators from a single time-delayed feedback loop, as described in detail in Ref.~\cite{hart2017experiments}. 
A schematic illustration of the experimental setup can be found in the Supplemental Material \cite{SM}.

\begin{figure}[t]
\centering
\includegraphics[width=1\columnwidth]{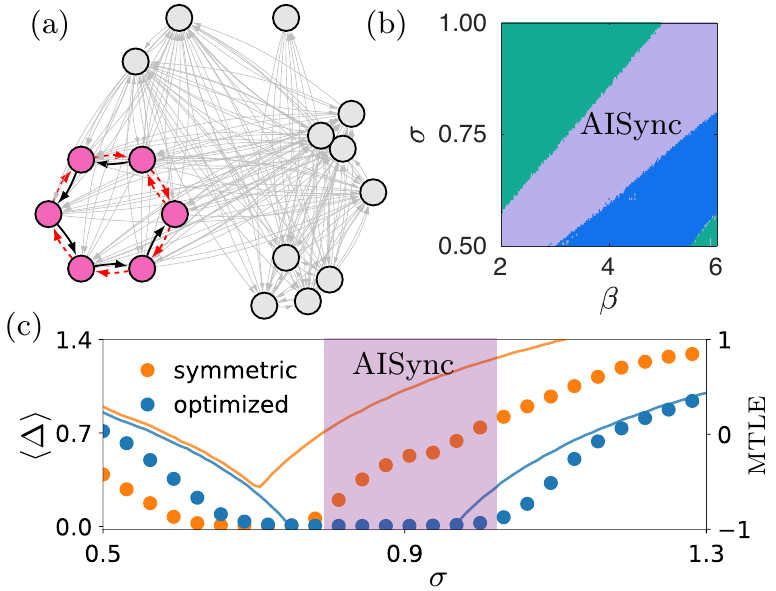}
\vspace{-5mm}
\caption{Experimental demonstration of structural AISync in a nonintertwined cluster. 
(a) Example network in which a symmetry cluster (magenta) is optimized for synchronization by removing the red links.
(b) Predictions based on the theoretical computation of the  MTLE, showing that in the $\sigma\times\beta$ parameter space there is an AISync region (purple); the other colors indicate the regions where both clusters synchronize (blue) and where neither cluster can synchronize (green).
(c) Experimentally measured average synchronization error $\langle\Delta\rangle$ in the original (orange) and optimized (blue) clusters for $\beta=6$.  
The experimental results are in good agreement with the MTLE calculations (color-coded curves).
}
\label{fig:4}
\end{figure}

We first consider the network configuration shown in \cref{fig:4}(a), which is a complex network with five symmetry clusters.
The symmetry cluster highlighted in magenta is nonintertwined, and can be optimized by removing the red dashed links. 
The MTLE calculation in \cref{fig:4}(b) predicts AISync to be common in the parameter space. 
Fixing $\beta = 6$, we performed $8$ runs of the experiment starting from different random initial conditions, and measured the normalized voltages $x_i$ for $8196$ iterations at each fixed coupling strength before increasing $\sigma$ by 0.015. 
The synchronization error is defined as $\Delta=\sqrt{\sum_{1\leq i\leq m}\|x_i-\bar{x}\|^2/m}$, 
where $\bar{x}$ is the mean inside the cluster.
%where $\| \, \|$ is the shortest distance between two points on a circle, and the mean $\bar{x}$ on a circle can be found as the phase of $\sum_{1\leq j\leq n}\mathrm{e}^{\mathrm{i}x_j}$. 
The data points in \cref{fig:4}(c) correspond to the average synchronization error $\langle\Delta\rangle$, defined as $\Delta$ averaged over the last $5000$ iterations for each $\sigma$ and then further averaged over the $8$ experimental runs. 
The error bars corresponding to the standard deviation across different runs are smaller than the size of the symbols.
%, and are not shown. 
One can observe AISync over a wide range of the coupling strength $\sigma$, matching the theoretical prediction.
Structural AISync is also common for different oscillator types and network structures and is robust against noise and parameter mismatches, as demonstrated systematically in the Supplemental Material \cite{SM}.

\begin{figure}[t]
\centering
\includegraphics[width=1\columnwidth]{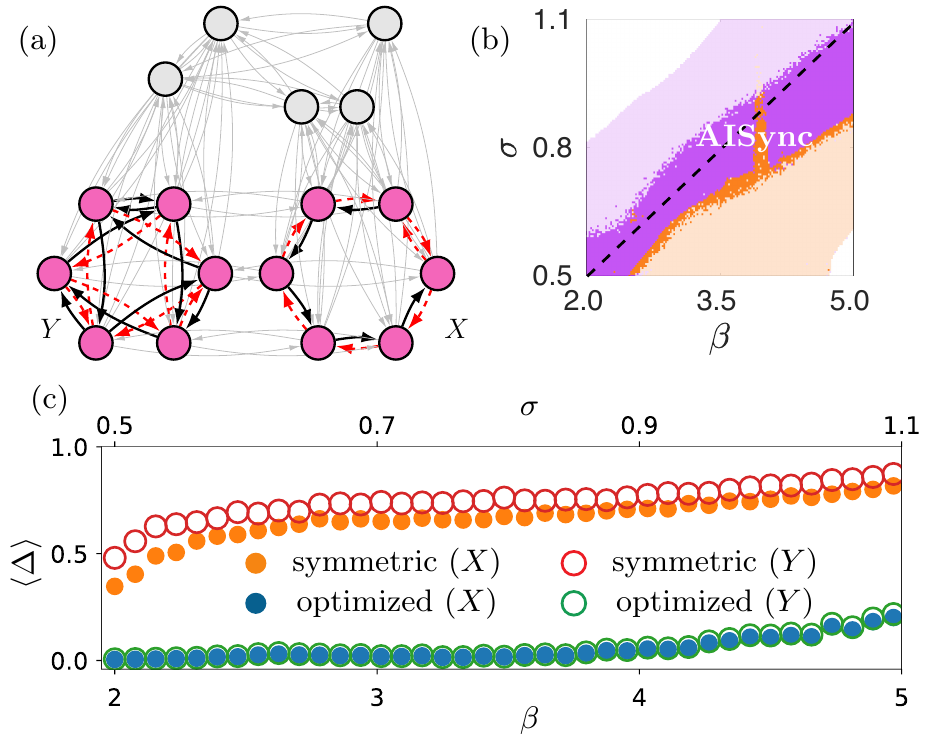}
\vspace{-5mm}
\caption{Demonstration of structural AISync in intertwined clusters. 
(a) Network in which two intertwined clusters (magenta) are optimized to induce synchronization by removing the red links.
(b) Region in the $\sigma\times\beta$ parameter space satisfying the condition in \cref{cdn:1}, which is expanded from the orange shaded area to include the purple shaded area when the clusters are optimized. The dark shades (orange and purple) highlight the AISync region determined through direct simulations.
(c) Experimentally measured average synchronization error $\langle\Delta\rangle$ in the original and optimized clusters when moving through the parameter space quasistatically along the dashed line in~(b).}
\label{fig:5}
\end{figure}

We now turn to the case of intertwined clusters. 
Consider two intertwined clusters $X$ and $Y$ subject to transverse perturbations $\delta\bm{X}$ and $\delta\bm{Y}$, respectively.
The variational equation for $\delta\bm{X}$ has the same form as \cref{eq:2} except for an additional cross-coupling term $\sigma\bm{C} \otimes J\bm{h}(\bm{s}_{I_Y}) \delta\bm{Y}$ added to the right, 
%where $\bm{C}_{ij} = 1$ if the $i$-th oscillator in cluster $X$ receives an input from the $j$-th oscillator in cluster $Y$ and $\bm{C}_{ij} = 0$ otherwise.
where $\bm{C}$ is the adjacency matrix describing the intercluster coupling from cluster $Y$ to cluster $X$. 
The variational equation for $\delta\bm{Y}$ is defined similarly.
Now, if $\delta\bm{X}$ ($\delta\bm{Y}$) does not converge to zero according to \cref{eq:2}, then the cross-coupling term must not vanish and $\|\delta\bm{Y}\|$ ($\|\delta\bm{X}\|$) must stay away from zero in order for $\|\delta\bm{X}\| \rightarrow 0$ ($\|\delta\bm{Y}\| \rightarrow 0$) in the full variational equation.
Thus, in order to stabilize synchronization in intertwined clusters, 
%(i.e., $\delta\bm{X}$ and $\delta\bm{Y}$ both going to zero), 
the following condition must be satisfied for each cluster:
%\begin{condition}
\begin{equation}
%\begin{split}
\|\bm{\eta}_i\| \rightarrow 0 \text{ in \cref{eq:3} for all transverse modes.}
\label{cdn:1}
%\end{split}
\end{equation}
%\end{condition}
In other words, $\|\delta\bm{X}\|$ and $\|\delta\bm{Y}\|$ converging to zero in \cref{eq:2} is a necessary condition for stable synchronization in $X$ and $Y$.
Because optimizing the clusters 
independently (as if they were nonintertwined)
%as if they were nonintertwined 
is guaranteed to expand the region satisfying the condition in \cref{cdn:1}, 
such independent optimization is an effective strategy for improving synchronization in intertwined clusters.
%the previous approach developed for nonintertwined clusters is also effective for improving the synchronization in intertwined clusters.
%Thus, the same optimization strategy, including all algorithms, can be easily extended to treat the case of intertwined clusters.
For more details on this analysis, see Supplemental Material~\cite{SM}.

We demonstrate the strength of our approach on a random network containing two intertwined clusters, which are highlighted in \cref{fig:5}(a).
Each cluster is optimized by removing the red dashed links, which breaks the structural symmetry but reduces the eigenratio of the cluster to 1.
The orange shade in \cref{fig:5}(b) indicates the region where the condition in \cref{cdn:1} is satisfied by the original clusters.
The region satisfying this condition is expanded to include the purple region when the clusters are optimized.
Direct simulations allow us to identify a large parameter region exhibiting AISync, which is highlighted in dark shades in \cref{fig:5}(b) and is included mainly in the expanded (purple) region.
A small portion of the AISync region also extends into the orange region, which follows from the condition in \cref{cdn:1} being necessary but not sufficient for synchronization in the original clusters.
%, thus not all parameters in the orange region support stable synchronization in the symmetry clusters.
%Not all points in the expanded region show structural AISync for the intertwined clusters, since the condition is necessary but not sufficient.
%Nonetheless, a significant portion of them do, which is verified by direct simulations and highlighted in dark \st{purple} {\YZZ shades}.
To validate the theory and the numerics, we perform experiments with parameters varied quasistatically along the dashed line in \cref{fig:5}(b).
%\st{, where the same procedure is used to obtain the average synchronization error $\langle\Delta\rangle$ of each cluster inside the AISync region.}
%It can be seen that the experiments agree well with the predictions in \cref{fig:5}(b).
As shown in \cref{fig:5}(c), the symmetry clusters are both incoherent for the entire range of parameters studied. 
The two optimized clusters exhibit perfectly synchronized dynamics except at the very edge of the AISync region, where the noise in the ADC has a marked impact on the dynamics (nevertheless, they are still much more synchronized than the symmetry clusters).
It is interesting to mention that although both optimized clusters are in synchrony themselves, they are not synchronized with each other.

In summary, we established the role of structural asymmetry (or structural heterogeneity) in promoting spontaneous synchronization through both theory and experiments. 
Our theory confirmed the generality of the phenomenon, while our experiments demonstrated its robustness.
Because symmetry clusters arise naturally in complex networks, 
our findings are applicable to a wide range of coupled dynamical systems. 
In particular, since identical synchronization in a symmetry cluster is the basic building block of more complex synchronization patterns, 
%we argue that 
our results can be used for the {\it targeted} topological control of cluster synchronization in complex networks, which echoes the positive effect of structural asymmetry on input control \cite{whalen2015observability}.

\vspace*{5mm}

%\begin{acknowledgments}
The authors thank Alex Mercanti, Takashi Nishikawa, Don Schmadel, and Thomas E. Murphy for insightful discussions.
This work was supported by ONR Grant No.\ N000141612481 (J.D.H. and R.R.) 
and ARO Grant No.\ W911NF-15-1-0272 (Y.Z. and A.E.M.).

J.D.H. and Y.Z. contributed equally to this work.
%\end{acknowledgments}

\let\oldaddcontentsline\addcontentsline% Store \addcontentsline
\renewcommand{\addcontentsline}[3]{}% Make \addcontentsline a no-op
\bibliographystyle{prl_title2}
\bibliography{net_dyn}

\begin{thebibliography}{10}
\providecommand{\bibnamefont}[1]{#1}
\providecommand{\bibfnamefont}[1]{#1}
\providecommand{\selectlanguage}[1]{\relax}

\bibitem{stewart2003symmetry}
\bibfnamefont{I.}~\bibnamefont{Stewart},
  \bibfnamefont{M.}~\bibnamefont{Golubitsky}, and
  \bibfnamefont{M.}~\bibnamefont{Pivato}, Symmetry groupoids and patterns of
  synchrony in coupled cell networks, SIAM J. Appl. Dyn. Syst. \textbf{2}, 609
  (2003).

\bibitem{nicosia2013remote}
\bibfnamefont{V.}~\bibnamefont{Nicosia},
  \bibfnamefont{M.}~\bibnamefont{Valencia},
  \bibfnamefont{M.}~\bibnamefont{Chavez},
  \bibfnamefont{A.}~\bibnamefont{D{\'\i}az-Guilera}, and
  \bibfnamefont{V.}~\bibnamefont{Latora}, Remote synchronization reveals
  network symmetries and functional modules, Phys. Rev. Lett. \textbf{110},
  174102 (2013).

\bibitem{aguiar2011dynamics}
\bibfnamefont{M.}~\bibnamefont{Aguiar}, \bibfnamefont{P.}~\bibnamefont{Ashwin},
  \bibfnamefont{A.}~\bibnamefont{Dias}, and
  \bibfnamefont{M.}~\bibnamefont{Field}, Dynamics of coupled cell networks:
  {S}ynchrony, heteroclinic cycles and inflation, J. Nonlinear Sci.
  \textbf{21}, 271 (2011).

\bibitem{pecora2014cluster}
\bibfnamefont{L.~M.} \bibnamefont{Pecora},
  \bibfnamefont{F.}~\bibnamefont{Sorrentino}, \bibfnamefont{A.~M.}
  \bibnamefont{Hagerstrom}, \bibfnamefont{T.~E.} \bibnamefont{Murphy}, and
  \bibfnamefont{R.}~\bibnamefont{Roy}, Cluster synchronization and isolated
  desynchronization in complex networks with symmetries, Nat. Commun.
  \textbf{5}, 4079 (2014).

\bibitem{donetti2005entangled}
\bibfnamefont{L.}~\bibnamefont{Donetti}, \bibfnamefont{P.~I.}
  \bibnamefont{Hurtado}, and \bibfnamefont{M.~A.} \bibnamefont{Munoz},
  Entangled networks, synchronization, and optimal network topology, Phys. Rev.
  Lett. \textbf{95}, 188701 (2005).

\bibitem{denker2004breaking}
\bibfnamefont{M.}~\bibnamefont{Denker}, \bibfnamefont{M.}~\bibnamefont{Timme},
  \bibfnamefont{M.}~\bibnamefont{Diesmann},
  \bibfnamefont{F.}~\bibnamefont{Wolf}, and
  \bibfnamefont{T.}~\bibnamefont{Geisel}, Breaking synchrony by heterogeneity
  in complex networks, Phys. Rev. Lett. \textbf{92}, 074103 (2004).

\bibitem{nishikawa2003heterogeneity}
\bibfnamefont{T.}~\bibnamefont{Nishikawa}, \bibfnamefont{A.~E.}
  \bibnamefont{Motter}, \bibfnamefont{Y.-C.} \bibnamefont{Lai}, and
  \bibfnamefont{F.~C.} \bibnamefont{Hoppensteadt}, Heterogeneity in oscillator
  networks: Are smaller worlds easier to synchronize?, Phys. Rev. Lett.
  \textbf{91}, 014101 (2003).

\bibitem{macarthur2008symmetry}
\bibfnamefont{B.~D.} \bibnamefont{MacArthur}, \bibfnamefont{R.~J.}
  \bibnamefont{S{\'a}nchez-Garc{\'\i}a}, and \bibfnamefont{J.~W.}
  \bibnamefont{Anderson}, Symmetry in complex networks, Discr. Appl. Math.
  \textbf{156}, 3525 (2008).

\bibitem{sorrentino2016complete}
\bibfnamefont{F.}~\bibnamefont{Sorrentino}, \bibfnamefont{L.~M.}
  \bibnamefont{Pecora}, \bibfnamefont{A.~M.} \bibnamefont{Hagerstrom},
  \bibfnamefont{T.~E.} \bibnamefont{Murphy}, and
  \bibfnamefont{R.}~\bibnamefont{Roy}, Complete characterization of the
  stability of cluster synchronization in complex dynamical networks, Sci. Adv.
  \textbf{2}, e1501737 (2016).

\bibitem{williams2013experimental}
\bibfnamefont{C.~R.} \bibnamefont{Williams}, \bibfnamefont{T.~E.}
  \bibnamefont{Murphy}, \bibfnamefont{R.}~\bibnamefont{Roy},
  \bibfnamefont{F.}~\bibnamefont{Sorrentino},
  \bibfnamefont{T.}~\bibnamefont{Dahms}, and
  \bibfnamefont{E.}~\bibnamefont{Sch{\"o}ll}, Experimental observations of
  group synchrony in a system of chaotic optoelectronic oscillators, Phys. Rev.
  Lett. \textbf{110}, 064104 (2013).

\bibitem{nixon2012controlling}
\bibfnamefont{M.}~\bibnamefont{Nixon}, \bibfnamefont{M.}~\bibnamefont{Fridman},
  \bibfnamefont{E.}~\bibnamefont{Ronen}, \bibfnamefont{A.~A.}
  \bibnamefont{Friesem}, \bibfnamefont{N.}~\bibnamefont{Davidson}, and
  \bibfnamefont{I.}~\bibnamefont{Kanter}, Controlling synchronization in large
  laser networks, Phys. Rev. Lett. \textbf{108}, 214101 (2012).

\bibitem{argyris2016experimental}
\bibfnamefont{A.}~\bibnamefont{Argyris},
  \bibfnamefont{M.}~\bibnamefont{Bourmpos}, and
  \bibfnamefont{D.}~\bibnamefont{Syvridis}, Experimental synchrony of
  semiconductor lasers in coupled networks, Opt. Exp. \textbf{24}, 5600 (2016).

\bibitem{rosin2013control}
\bibfnamefont{D.~P.} \bibnamefont{Rosin},
  \bibfnamefont{D.}~\bibnamefont{Rontani}, \bibfnamefont{D.~J.}
  \bibnamefont{Gauthier}, and \bibfnamefont{E.}~\bibnamefont{Sch{\"o}ll},
  Control of synchronization patterns in neural-like Boolean networks, Phys.
  Rev. Lett. \textbf{110}, 104102 (2013).

\bibitem{vardi2012synchronization}
\bibfnamefont{R.}~\bibnamefont{Vardi}, \bibfnamefont{R.}~\bibnamefont{Timor},
  \bibfnamefont{S.}~\bibnamefont{Marom},
  \bibfnamefont{M.}~\bibnamefont{Abeles}, and
  \bibfnamefont{I.}~\bibnamefont{Kanter}, Synchronization with mismatched
  synaptic delays: {A} unique role of elastic neuronal latency, Europhys. Lett.
  \textbf{100}, 48003 (2012).

\bibitem{takamatsu2001spatiotemporal}
\bibfnamefont{A.}~\bibnamefont{Takamatsu},
  \bibfnamefont{R.}~\bibnamefont{Tanaka},
  \bibfnamefont{H.}~\bibnamefont{Yamada},
  \bibfnamefont{T.}~\bibnamefont{Nakagaki},
  \bibfnamefont{T.}~\bibnamefont{Fujii}, and
  \bibfnamefont{I.}~\bibnamefont{Endo}, Spatiotemporal symmetry in rings of
  coupled biological oscillators of Physarum plasmodial slime mold, Phys. Rev.
  Lett. \textbf{87}, 078102 (2001).

\bibitem{totz2015phase}
\bibfnamefont{J.~F.} \bibnamefont{Totz}, \bibfnamefont{R.}~\bibnamefont{Snari},
  \bibfnamefont{D.}~\bibnamefont{Yengi}, \bibfnamefont{M.~R.}
  \bibnamefont{Tinsley}, \bibfnamefont{H.}~\bibnamefont{Engel}, and
  \bibfnamefont{K.}~\bibnamefont{Showalter}, Phase-lag synchronization in
  networks of coupled chemical oscillators, Phys. Rev. E \textbf{92}, 022819
  (2015).

\bibitem{hart2016experimental}
\bibfnamefont{J.~D.} \bibnamefont{Hart},
  \bibfnamefont{K.}~\bibnamefont{Bansal}, \bibfnamefont{T.~E.}
  \bibnamefont{Murphy}, and \bibfnamefont{R.}~\bibnamefont{Roy}, Experimental
  observation of chimera and cluster states in a minimal globally coupled
  network, Chaos \textbf{26}, 094801 (2016).

\bibitem{PhysRevLett.117.114101}
\bibfnamefont{T.}~\bibnamefont{Nishikawa} and \bibfnamefont{A.~E.}
  \bibnamefont{Motter}, Symmetric States Requiring System Asymmetry, Phys. Rev.
  Lett. \textbf{117}, 114101 (2016).

\bibitem{zhang2017asymmetry}
\bibfnamefont{Y.}~\bibnamefont{Zhang},
  \bibfnamefont{T.}~\bibnamefont{Nishikawa}, and \bibfnamefont{A.~E.}
  \bibnamefont{Motter}, Asymmetry-induced synchronization in oscillator
  networks, Phys. Rev. E \textbf{95}, 062215 (2017).

\bibitem{zhang2017nonlinearity}
\bibfnamefont{Y.}~\bibnamefont{Zhang} and \bibfnamefont{A.~E.}
  \bibnamefont{Motter}, Identical synchronization of nonidentical oscillators:
  When only birds of different feathers flock together, Nonlinearity
  \textbf{31}, R1 (2018).

\bibitem{cho2017stable}
\bibfnamefont{Y.~S.} \bibnamefont{Cho},
  \bibfnamefont{T.}~\bibnamefont{Nishikawa}, and \bibfnamefont{A.~E.}
  \bibnamefont{Motter}, Stable Chimeras and Independently Synchronizable
  Clusters, Phys. Rev. Lett. \textbf{119}, 084101 (2017).

\bibitem{pecora1998master}
\bibfnamefont{L.~M.} \bibnamefont{Pecora} and \bibfnamefont{T.~L.}
  \bibnamefont{Carroll}, Master stability functions for synchronized coupled
  systems, Phys. Rev. Lett. \textbf{80}, 2109 (1998).

\bibitem{nishikawa2006maximum}
\bibfnamefont{T.}~\bibnamefont{Nishikawa} and \bibfnamefont{A.~E.}
  \bibnamefont{Motter}, Maximum performance at minimum cost in network
  synchronization, Physica (Amsterdam) \textbf{224D}, 77 (2006).

\bibitem{SM}
See Supplemental Material for details of the experimental system, stability
  analysis, optimization algorithm, and additional examples of AISync.

\bibitem{barahona2002synchronization}
\bibfnamefont{M.}~\bibnamefont{Barahona} and \bibfnamefont{L.~M.}
  \bibnamefont{Pecora}, Synchronization in small-world systems, Phys. Rev.
  Lett. \textbf{89}, 054101 (2002).

\bibitem{li2010consensus}
\bibfnamefont{Z.}~\bibnamefont{Li}, \bibfnamefont{Z.}~\bibnamefont{Duan},
  \bibfnamefont{G.}~\bibnamefont{Chen}, and
  \bibfnamefont{L.}~\bibnamefont{Huang}, Consensus of multiagent systems and
  synchronization of complex networks: A unified viewpoint, IEEE Trans.
  Circuits Syst. I \textbf{57}, 213 (2010).

\bibitem{flunkert2010synchronizing}
\bibfnamefont{V.}~\bibnamefont{Flunkert},
  \bibfnamefont{S.}~\bibnamefont{Yanchuk},
  \bibfnamefont{T.}~\bibnamefont{Dahms}, and
  \bibfnamefont{E.}~\bibnamefont{Sch{\"o}ll}, Synchronizing distant nodes: A
  universal classification of networks, Phys. Rev. Lett. \textbf{105}, 254101
  (2010).

\bibitem{huang2009generic}
\bibfnamefont{L.}~\bibnamefont{Huang}, \bibfnamefont{Q.}~\bibnamefont{Chen},
  \bibfnamefont{Y.-C.} \bibnamefont{Lai}, and \bibfnamefont{L.~M.}
  \bibnamefont{Pecora}, Generic behavior of master-stability functions in
  coupled nonlinear dynamical systems, Phys. Rev. E \textbf{80}, 036204 (2009).

\bibitem{comment1}
For nonlinear oscillators, this can be done numerically by calculating the
  master stability function for a sufficiently large region in the complex
  plane that encompasses all eigenvalues of the coupling matrix scaled by the
  permissible coupling strength.

\bibitem{nishikawa2010network}
\bibfnamefont{T.}~\bibnamefont{Nishikawa} and \bibfnamefont{A.~E.}
  \bibnamefont{Motter}, Network synchronization landscape reveals compensatory
  structures, quantization, and the positive effect of negative interactions,
  Proc. Natl. Acad. Sci. U.S.A. \textbf{107}, 10342 (2010).

\bibitem{comment}
Note that although structural symmetry is broken in this process, the
  oscillators can still synchronize identically as a Laplacian cluster [9]
  because of the diffusive coupling.

\bibitem{biggs1993algebraic}
\bibfnamefont{N.}~\bibnamefont{Biggs}, \emph{Algebraic Graph Theory} (Cambridge
  University Press, Cambridge, England, 1993).

\bibitem{mckay2014practical}
\bibfnamefont{B.~D.} \bibnamefont{McKay} and
  \bibfnamefont{A.}~\bibnamefont{Piperno}, Practical graph isomorphism, II, J.
  Symb. Comput. \textbf{60}, 94 (2014).

\bibitem{nishikawa2006synchronization}
\bibfnamefont{T.}~\bibnamefont{Nishikawa} and \bibfnamefont{A.~E.}
  \bibnamefont{Motter}, Synchronization is optimal in nondiagonalizable
  networks, Phys. Rev. E \textbf{73}, 065106 (2006).

\bibitem{SA}
Simulated annealing code to improve synchronizability through minimal link
  rewiring, removal, or addition:
  \url{https://github.com/y-z-zhang/optimize_sym_cluster/}.

\bibitem{hart2017experiments}
\bibfnamefont{J.~D.} \bibnamefont{Hart}, \bibfnamefont{D.~C.}
  \bibnamefont{Schmadel}, \bibfnamefont{T.~E.} \bibnamefont{Murphy}, and
  \bibfnamefont{R.}~\bibnamefont{Roy}, Experiments with arbitrary networks in
  time-multiplexed delay systems, Chaos \textbf{27}, 121103 (2017).

\bibitem{whalen2015observability}
\bibfnamefont{A.~J.} \bibnamefont{Whalen}, \bibfnamefont{S.~N.}
  \bibnamefont{Brennan}, \bibfnamefont{T.~D.} \bibnamefont{Sauer}, and
  \bibfnamefont{S.~J.} \bibnamefont{Schiff}, Observability and controllability
  of nonlinear networks: The role of symmetry, Phys. Rev. X \textbf{5}, 011005
  (2015).

\bibitem{godsil1982eigenvalues}
\bibfnamefont{C.}~\bibnamefont{Godsil}, Eigenvalues of graphs and digraphs,
  Linear Algebra Appl. \textbf{46}, 43 (1982).

\end{thebibliography}
\let\addcontentsline\oldaddcontentsline% Restore \addcontentsline

%\end{document}

%%%%%%%%%%%%%%%%
%%%%%%%%%%%%%%%%

\clearpage
\onecolumngrid
\setcounter{page}{1}
\renewcommand{\thepage}{S\arabic{page}}
\setcounter{equation}{0}
\renewcommand{\theequation}{S\arabic{equation}}
\setcounter{figure}{0}
\renewcommand{\thefigure}{S\arabic{figure}}
\setcounter{section}{0}
\renewcommand{\thesection}{S\arabic{section}}
\setcounter{table}{0}
\renewcommand{\thetable}{S\arabic{table}}

\begin{center}
{\Large\bf Supplemental Material}\\[3mm]
{\large{\mytitle}}\\[1pt]
Joseph D. Hart, Yuanzhao Zhang, Rajarshi Roy, and Adilson E. Motter
\end{center}

\tableofcontents

\section{Experimental implementation of the coupled optoelectronic oscillators}

\begin{figure}[ht]
\centering
\subfloat[]{
  \includegraphics[width=.45\textwidth]{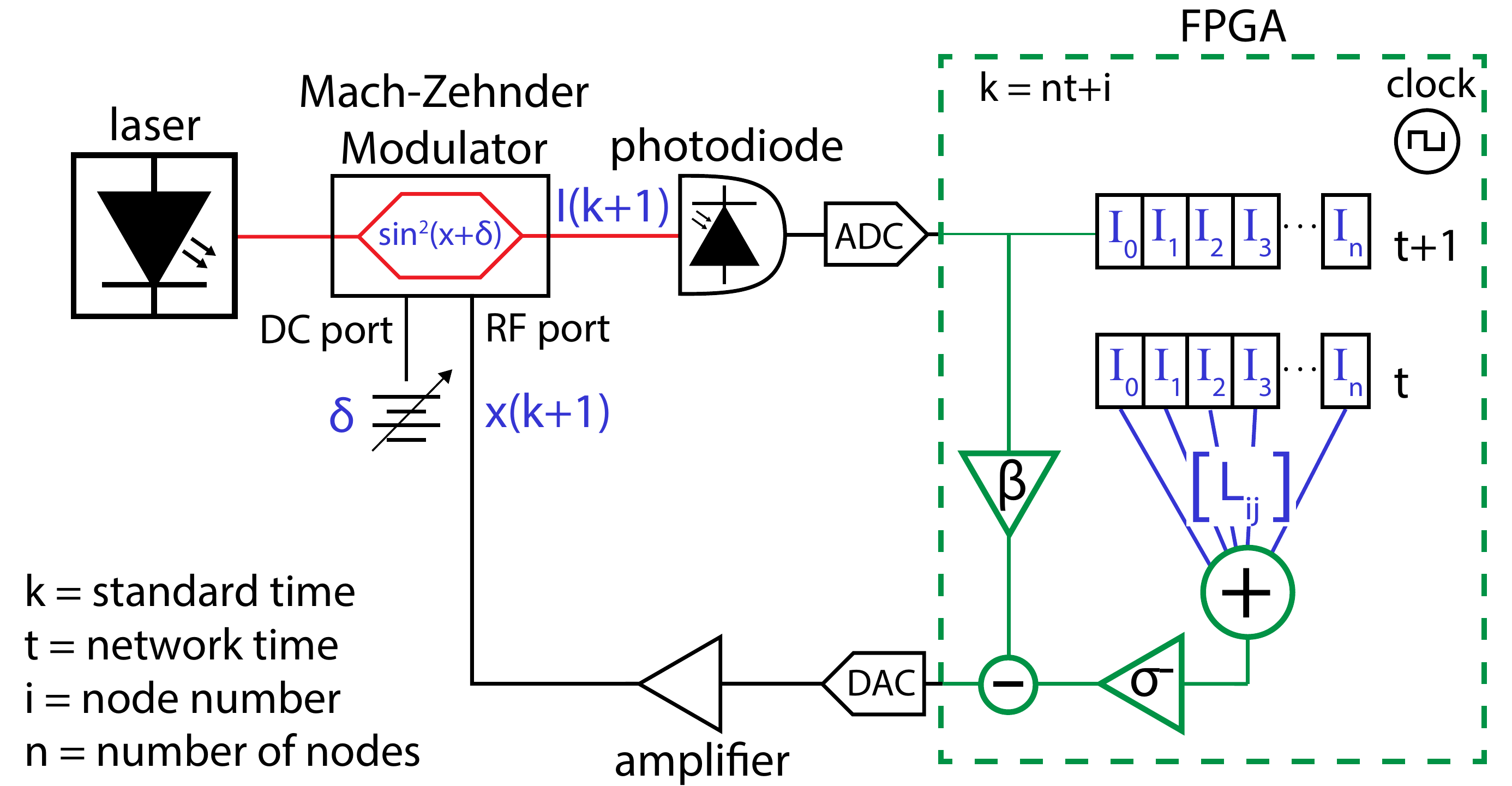}
}
\vspace{-5mm}
\caption{Schematic illustration of the 
apparatus used in our experiments. 
The diagram shows the components of one optoelectronic oscillator (left) and associated coupling scheme (right), 
which is implemented using time multiplexing in the FPGA.
}
\label{fig:2}
\end{figure}

\newpage

\section{Optimizing intertwined clusters}

In this section, we provide more details on the optimization of intertwined clusters. 
When two clusters are intertwined, desynchronization in one cluster will in general lead to the loss of synchrony in the other cluster (an example would be two equal-sized rings coupled in one-to-one fashion).
This is because the symmetry group acting on the two clusters does not admit a geometric decomposition; that is, symmetry permutations cannot be applied to each cluster independently. 
As a consequence, a desynchronized cluster sends incoherent signals to nodes in the other cluster, causing its intertwined counterpart to desynchronize as well.
The irreducible representation transformation introduced in 
Ref.~\cite{pecora2014cluster} 
%Ref.~[4]
is a powerful formalism that enables stability analysis on many cluster synchronization patterns.
In that framework, the presence of intertwined clusters is reflected in nontrivial transverse blocks (i.e., blocks with dimension greater than 1) in the transformed coupling matrix, whereas nonintertwined clusters only give rise to $1 \times 1$ transverse blocks.
Unfortunately, the high dimensionality of the transverse blocks makes the effect of topological perturbations on cluster synchronizability opaque, and thus the analysis of the transformed matrix offers little insight into how to optimize the clusters to support desired synchronization patterns.

We developed a new perspective that gives a simple necessary condition for the synchronization in intertwined clusters.
This in turn points to an extension of the previous optimization scheme that is no longer limited to nonintertwined clusters.

Consider two intertwined clusters $X$ and $Y$ subject to transverse perturbations $\delta\bm{X}$ and $\delta\bm{Y}$, respectively.
Their variational equation reads
\begin{equation}
  \begin{cases}
    \delta\dot{\bm{X}} = \left( \mathds{1}_{m} \otimes J\bm{f}(\bm{s}_{I_X}) - \sigma \widehat{\bm{L}}_X \otimes J\bm{h}(\bm{s}_{I_X}) \right) \delta\bm{X} + \sigma\bm{C} \otimes J\bm{h}(\bm{s}_{I_Y}) \delta\bm{Y}, \\
    \delta\dot{\bm{Y}} = \left( \mathds{1}_{m'} \otimes J\bm{f}(\bm{s}_{I_Y}) - \sigma \widehat{\bm{L}}_Y \otimes J\bm{h}(\bm{s}_{I_Y}) \right) \delta\bm{Y} + \sigma\bm{D} \otimes J\bm{h}(\bm{s}_{I_X}) \delta\bm{X}.
  \end{cases}
  \label{eq:s7}
\end{equation}
Here, $\bm{C}_{ij} = 1$ if the $i$-th oscillator in cluster $X$ receives an input from the $j$-th oscillator in cluster $Y$ and $\bm{C}_{ij} = 0$ otherwise. 
The intercluster coupling matrix $\bm{D}$ is similarly defined with the role of two clusters exchanged ($\bm{D} = \bm{C}^\intercal$ if the intercluster coupling is undirected).
Without the cross-coupling term, \cref{eq:s7} reduces to the nonintertwined case discussed in the main text
\begin{equation}
  \begin{cases}
    \delta\dot{\bm{X}} = \left( \mathds{1}_{m} \otimes J\bm{f}(\bm{s}_{I_X}) - \sigma \widehat{\bm{L}}_X \otimes J\bm{h}(\bm{s}_{I_X}) \right) \delta\bm{X}, \\
    \delta\dot{\bm{Y}} = \left( \mathds{1}_{m'} \otimes J\bm{f}(\bm{s}_{I_Y}) - \sigma \widehat{\bm{L}}_Y \otimes J\bm{h}(\bm{s}_{I_Y}) \right) \delta\bm{Y}.
  \end{cases}
  \label{eq:s8}
\end{equation}
Because of the intertwined nature of the two clusters, they must be considered concurrently when synchronization is desired in either of them.
That is, $\widehat{\bm{L}}_X$ and $\widehat{\bm{L}}_Y$ should be optimized to ensure that $\delta\bm{X}$ and $\delta\bm{Y}$ both vanish in \cref{eq:s7}.

It is difficult to establish a synchronizability measure on two clusters based on \cref{eq:s7}, but we can see the following connection between \cref{eq:s7,eq:s8}:
\begin{equation}
  \|\delta\bm{X}\|\rightarrow 0 \text{ and } \|\delta\bm{Y}\|\rightarrow 0 \text{ in \cref{eq:s7}}
\end{equation}
\hspace{8cm}$\Downarrow$
\begin{equation}
  \|\delta\bm{X}\|\rightarrow 0 \text{ and } \|\delta\bm{Y}\|\rightarrow 0 \text{ in \cref{eq:s8}}.
  \label{eq:s9}
\end{equation}
%\hspace{-2mm}
That is, $\|\delta\bm{X}\|$ and $\|\delta\bm{Y}\|$ going to zero in \cref{eq:s8} is a necessary condition for the synchronization in intertwined clusters.
For example, if $\|\delta\bm{X}\|$ does not vanish in \cref{eq:s8}, then $\|\delta\bm{Y}\|$ must be away from zero in order for $\|\delta\bm{X}\|\rightarrow 0$ in \cref{eq:s7}.
This connection between \cref{eq:s7,eq:s8} implies that we can promote synchronization in the intertwined clusters by optimizing each of the two clusters independently, using the same method originally developed for nonintertwined clusters. 
In particular, such optimization is guaranteed to expand the region in parameter space satisfying the necessary condition in \cref{eq:s9} (i.e., the condition in \cref{cdn:1} in the main text).
Inside this expanded region, one is likely to observe structural AISync, as experimentally demonstrated in the main text.
It is worth mentioning that the same argument still holds when more than two clusters are intertwined.

\newpage

\section{Stability analysis of nondiagonalizable clusters}

When dealing with directed networks, one must be aware of the possibility of nondiagonalizable coupling matrices, which can be the case even for symmetric networks 
\cite{godsil1982eigenvalues}.
%[25].
Here, we present details of how the  analysis in the manuscript also applies to nondiagonalizable networks. To demonstrate that, our key observation is that the treatment of nondiagonalizable networks in 
Refs.~\cite{nishikawa2006synchronization,nishikawa2006maximum}
%Refs.~[23,35] 
can be generalized to the case of a cluster subnetwork in which each oscillator receives a common input from the rest of the network.

We start from the variational equation of the system in the form of \cref{eq:2} in the main text,
\begin{equation}
  \delta\dot{\bm{X}} = \left[ \mathds{1}_m \otimes J\bm{f}(\bm{s}_I) - \sigma \widehat{\bm{L}} \otimes J\bm{h}(\bm{s}_I) \right] \delta\bm{X},
  \label{eq:s1}
\end{equation}
but this time we lift the assumption that the matrix $\widehat{\bm{L}}$ is diagonalizable.
For such systems, in general we can not find $m$ independent eigenvectors for $\widehat{\bm{L}}$.
Nevertheless, this matrix can always be transformed into a Jordan canonical form through a similarity transformation
defined by an invertible matrix $\bm{P}$, such that
\begin{equation}
  \bm{B} = \bm{P}^{-1}\widehat{\bm{L}}\bm{P} =  
  \begin{pmatrix}
    \widetilde{\mu} &  &  &  \\
     & \bm{B}_1 &  &  \\
     &  & \ddots & \\
     &  &  & \bm{B}_q 
  \end{pmatrix},\quad
  \bm{B}_j = 
  \begin{pmatrix}
     \widehat{v}_{j+1} &  &  &  \\
     1 & \widehat{v}_{j+1} &  &  \\
     & \ddots & \ddots & \\
     &  & 1 & \widehat{v}_{j+1} 
  \end{pmatrix},
\end{equation}
where $\widehat{v}_{j+1}$ is the eigenvalue of $\widehat{\bm{L}}$ corresponding to the Jordan block $\bm{B}_j$, and the matrix entries not shown are zero.
The eigenvalues are numbered from $2$ to $q+1$ for consistency with the eigenvalue notation in the main text, and are thus ordered as in the rest of the paper 
but now without relabeling the (identical) eigenvalues associated with the same Jordan block (which is why $q+1<m$ in the nondiagonalizable case).
The special case in which $\widehat{\bm{L}}$ is diagonalizable is also included in this transformation, and it merely corresponds to the case in which all Jordan blocks are $1\times 1$.

\Cref{eq:s1} can now be decoupled into $q+1$ independent equations accounting for the Jordan blocks.
The central difference between the case of an isolated network, as considered in 
Refs.~\cite{nishikawa2006synchronization,nishikawa2006maximum},
%Refs.~[23,35], 
and the cluster subnetworks considered here is the entry $B_{11} = \widetilde{\mu}$, which is zero for isolated networks. However, this term corresponds to 
a perturbation mode parallel to the cluster synchronization manifold and hence has no influence on the stability of the synchronization state. (The input connections 
from the rest of the network to the cluster also impact the synchronization state $\bm{s}_I$ and shift the eigenvalues $\widehat{v}_j$, but
those are not material differences since the same also occurs in the diagonalizable case.) Thus, to analyze the transverse modes, we focus on the $q$ block-decoupled equations associated with the Jordan blocks $\bm{B}_1, \cdots , \bm{B}_q$:
\begin{equation}
  \dot{\bm{\eta}}^{(j)} = \left[ \mathds{1}_{k} \otimes J\bm{f}(\bm{s}_I) - \sigma \bm{B}_j \otimes J\bm{h}(\bm{s}_I) \right] \bm{\eta}^{(j)}, \,\,\,\,\, j=1,\dots, q.
  \label{eq:s2}
\end{equation}

Assuming that $\bm{B}_j$ is $k \times k$,  the corresponding equation can be written explicitly for each mode as
\begin{equation}
  \begin{split}
  \dot{\bm{\eta}}^{(j)}_1 = & \left[ J\bm{f}(\bm{s}_I) - \sigma \widehat{v}_{j+1} J\bm{h}(\bm{s}_I) \right] \bm{\eta}^{(j)}_1, \\
  \dot{\bm{\eta}}^{(j)}_2 = & \left[ J\bm{f}(\bm{s}_I) - \sigma \widehat{v}_{j+1} J\bm{h}(\bm{s}_I) \right] \bm{\eta}^{(j)}_2 - \sigma J\bm{h}(\bm{s}_I) \bm{\eta}^{(j)}_1, \\
  \cdots \\
  \dot{\bm{\eta}}^{(j)}_k = & \left[ J\bm{f}(\bm{s}_I) - \sigma \widehat{v}_{j+1} J\bm{h}(\bm{s}_I) \right] \bm{\eta}^{(j)}_k - \sigma J\bm{h}(\bm{s}_I) \bm{\eta}^{(j)}_{k-1}. 
  \end{split}
  \label{eq:s3}
\end{equation}
Starting from the first equation in \cref{eq:s3}, we notice that $\bm{\eta}^{(j)}_1$ does not depend on any other $\bm{\eta}^{(j)}_i$ and its equation is exactly the master stability equation [\cref{eq:3} in the main text]. 
If \cref{eq:3} is stable for $\widehat{v}_{j+1}$, then $\bm{\eta}^{(j)}_1$ converges to zero exponentially.
Turning to the second equation in \cref{eq:s3}, we  can see that the influence of $\bm{\eta}^{(j)}_1$ on $\bm{\eta}^{(j)}_2$ vanishes and $\bm{\eta}^{(j)}_2$ will also approach zero as $t \rightarrow \infty$ (under the reasonable assumption that  $J\bm{h}(\bm{s}_I)$ is bounded).
Applying the same argument iteratively, it follows that the stability of \cref{eq:s3} is entirely determined by the stability of \cref{eq:3} for the eigenvalue $\widehat{v}_{j+1}$ (with $ \bm{\eta}_i$ demoted by $ \bm{\eta}^{(j)}_1$). 
The same applies for all $j$ and leads to the conclusion that, even if $\widehat{\bm{L}}$ is nondiagonalizable, the condition for the cluster synchronous state to be stable is that  $\Lambda (\sigma \widehat{v}_{j+1})<0$ for $j=1,\dots, q$, where $\Lambda$ is the largest Lyapunov exponent of \cref{eq:3} and  $\widehat{v}_2, \cdots, \widehat{v}_{q+1}$ represent the eigenvalues associated with the transverse modes. 
Therefore, our analysis of synchronizability presented in the main text (including the use of the eigenratio $R$) applies equally well to nondiagonalizable networks. 
%Depending on the shape of the stability region, synchronizability can in principle be influenced by other structural measures besides eigenratio, but in practice a significant reduction of eigenratio are usually guaranteed to improve synchronizability (as exemplified in \cref{fig:3}). 
%including the use of the eigenratio $R = \text{Re}(\widetilde{v}_{q+1})/\text{Re}(\widetilde{v}_2)$ to measure synchronizability. 
%In some specific cases, synchronizability may depend on the imaginary part of the eigenvalues, but this does not influence our results since the optimization of symmetry clusters necessarily results in real eigenvalues.

\newpage

\section{Improving synchronizability through minimal link rewiring}

\begin{figure}[h]
\centering
\includegraphics[width=.65\textwidth]{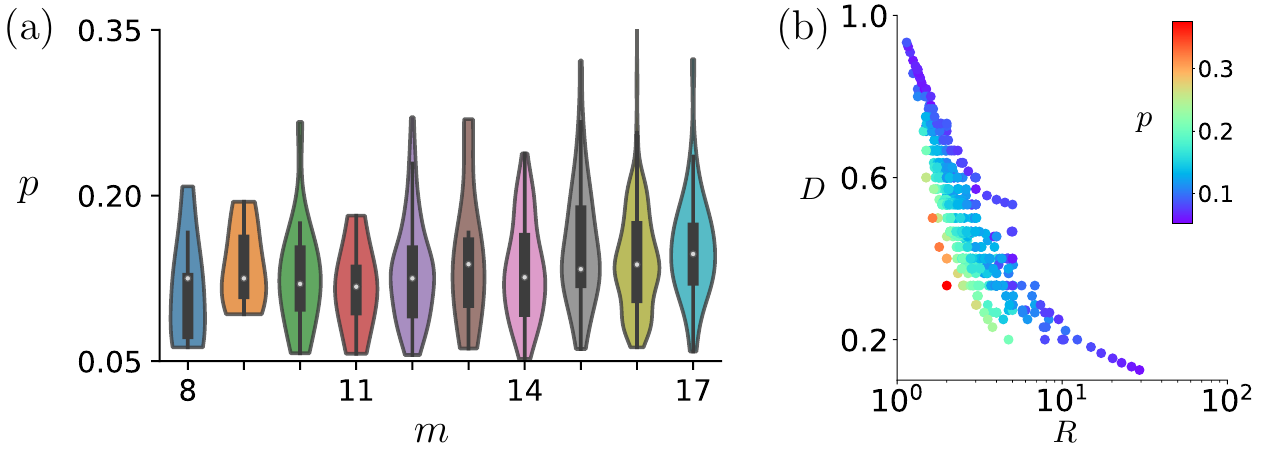}
%\vspace{-6mm}
\caption{Improvement of synchronizability by breaking the cluster symmetry through link rewiring. 
(a) Percentage of rewiring, $p$, needed to reduce the eigenratio gap $R-1$ by half. 
The violin plots show the kernel density estimation of $p$ over all connected undirected symmetry clusters for each cluster size $m$.
Inside each violin plot, the white dot represents the median of the data, the thick line indicates the interquartile range, and the thin line encompasses the 95\% confidence interval.
(b) Color-coded $p$ in the diagram of link density $D$ versus eigenratio $R$ for all symmetry clusters
considered in panel (a), where each cluster corresponds to one data point.}
\label{fig:s0}
\end{figure}

In this section, we consider the optimization of symmetry clusters by rewiring a small number of links. 
One rewiring consists of removing an existing link and adding a different link not yet present in the cluster.
Specifically, we developed an algorithm to optimize synchronizability by rewiring intra-cluster connections 
\cite{SA}, 
%[36],
which preserves the nonintertwined nature of the clusters.
This allows us to investigate how many directional links need to be rewired to reduce the eigenratio gap $R-1$ by half.
\Cref{fig:s0} summarizes results for all connected symmetry clusters that are undirected of sizes between $m=8$ and $17$, where the rewiring percentage $p = h/\ell$ is the ratio between the minimal number of link rewiring $h$ that halves $R-1$ and the total number $\ell$ of internal directed links of the cluster.
\Cref{fig:s0}(a) shows that on average only about $14\%$ of the links need to be rewired to significantly improve synchronizability of symmetry clusters, and it is largely size independent. 
{\YZ 
Our algorithm also works for link addition and link removal.
In the case of link addition, link density needs to increase by about $20\%$ on average to reduce the eigenratio gap to half;
for link removal, about $40\%$ of the links need to be removed to achieve the same effect.
}

\Cref{fig:s0}(b) shows the rewiring percentage $p$ as function of the eigenratio $R$ and link density $D = \frac{\ell}{m(m-1)}$, where each data point represents one symmetry cluster.
It is clear that clusters that are small in both $D$ and $R$ require the highest percentage of links to be rewired in order to significantly reduce the eigenratio gap.
This confirms the intuition that if a network achieves a small eigenratio with a relatively small number of links, then its organization is efficient and its synchronizability is relatively hard to improve.
Conversely, a dense non-optimal network or a network with a relatively large eigenratio is easy to optimize with a small number of link modifications.

\newpage

\section{Application of the minimal-rewiring algorithm to global synchronization}

\begin{figure}[htb]
\centering
\includegraphics[width=.5\columnwidth]{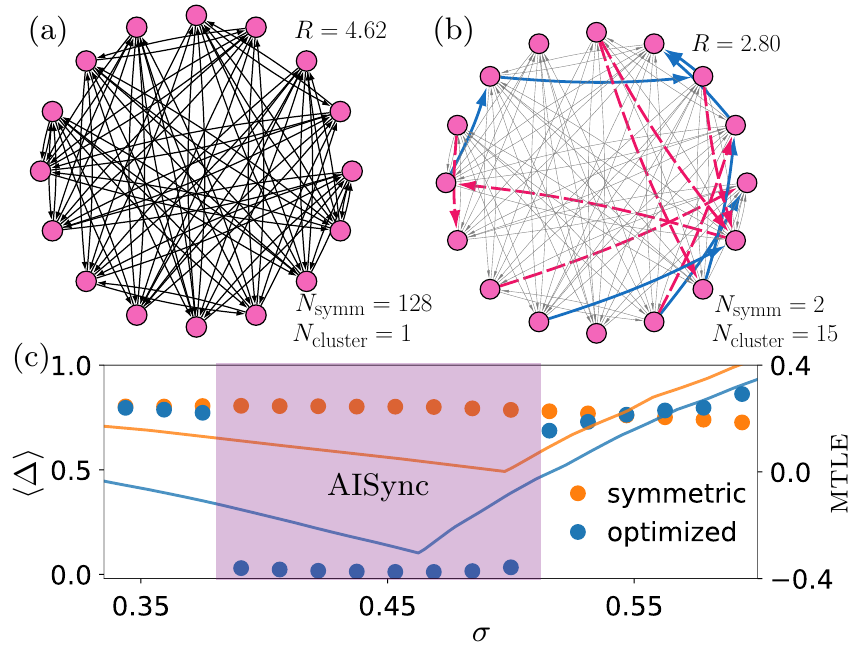}
\vspace{-5mm}
\caption{Experimental demonstration of structural AISync in global synchronization.
(a) Symmetric 16-node network of coupled optoelectronic oscillators used in the experiment, which has   $128$ symmetries, consists of a single (global) symmetry cluster, and has an eigenratio of $R=4.62$.
(b) Optimized network found through simulated annealing, where $7$ links are removed (red) and $7$ links are added (blue) to the network in (a), resulting in a network with only $2$ symmetries and $15$ symmetry clusters but an eigenratio of $R=2.80$.
(c) Experimentally measured average synchronization error $\langle\Delta\rangle$ for the symmetric and symmetry-broken network in (a) and (b), respectively, where only the latter can synchronize for a range of $\sigma$ (purple region). The feedback strength was set to $\beta=2.8$. This experimental result is consistent with the theoretically computed MTLE (color-coded curves), which is more negative for the symmetry-broken network. 
}
\label{fig:3}
\end{figure}

In this section, we apply the algorithm from the last section to a case in which the full network is symmetric and we seek to optimize global synchronization.
In \cref{fig:3} we study a $16$-node symmetric network and show explicitly through our experiments that it becomes more synchronizable with less symmetry. 
In the original network [\cref{fig:3}(a)], all nodes play exactly the same structural role. 
After seven directional link rewiring [marked in \cref{fig:3}(b)], the symmetry of the network is largely broken and almost all nodes are now structurally different: the original 16-node symmetry cluster is reduced to $14$ single-node clusters and only $2$ nodes occupying symmetric positions.
The eigenratio, however, reduces from $R=4.62$ to $R=2.80$ and thus improves significantly.

The experimental results are presented in \cref{fig:3}(c), where  we show the average synchronization error as a function of the coupling strength for both networks.
The experimental data clearly demonstrates that synchronization is only achieved for the network with reduced symmetry. 
The experimental result is consistent with the 
MTLE determined from numerical calculations of the variational equation of the model in \cref{eq:exp} [color-coded curves in \cref{fig:3}(c)]. 
Indeed, for values of $\sigma$ close to the boundary of linear stability, synchronization is not observed in experiments due to noise in the ADC 
\cite{hart2017experiments}, 
%[37],
but synchronization is consistently observed once  the MTLE becomes sufficiently negative.

\newpage

\section{Prevalence of structural \texorpdfstring{AIS\MakeLowercase{ync}}{AISync}}

\begin{figure}[!hbt]
\centering
\subfloat[]{
\resizebox{.23\textwidth}{!}{
    \begin{tikzpicture}[
      vertex/.style={draw,circle,very thick,minimum size=.6cm},
      arc/.style={draw,very thick,
      -{Latex[length=3mm, width=2mm]},
      bend left=10},
      arc1/.style={draw,thin,gray!50,
      -{Latex[length=2mm, width=1mm]},
      bend left=10}]
      \node[vertex,fill=magenta!60] (p12) at (.7,1.25) {};
      \node[vertex,fill=magenta!60] (p14) at (-1.45,0) {};
      \node[vertex,fill=magenta!60] (p16) at (.7,-1.25) {};
      \node[vertex,fill=magenta!60] (p13) at (-.7,1.25) {};
      \node[vertex,fill=magenta!60] (p15) at (-.7,-1.25) {};
      \node[vertex,fill=magenta!60] (p17) at (1.45,0) {}; 

      \node[vertex,fill=gray!20] (p1) at (5.5,.5) {};
      \node[vertex,fill=gray!20] (p2) at (4.8,1.5) {};
      \node[vertex,fill=gray!20] (p3) at (5.2,2.3) {};
      \node[vertex,fill=gray!20] (p9) at (4.1,1.8) {}; 

      \node[vertex,fill=gray!20] (p6) at (4.5,-1) {};
      \node[vertex,fill=gray!20] (p8) at (4,-1.5) {};
      \node[vertex,fill=gray!20] (p10) at (3,-1.5) {};

      \node[vertex,fill=gray!20] (p4) at (0,4.5) {};
      \node[vertex,fill=gray!20] (p5) at (.55,3.5) {};

      \node[vertex,fill=gray!20] (p7) at (4,4.5) {};
      \node[vertex,fill=gray!20] (p11) at (3,4) {};

      \begin{pgfonlayer}{bg}
      \foreach [count=\r] \row in 
      {{0,1,1,1,0,1,1,1,0,1,1,0,0,0,0,0,0},
       {1,0,0,0,1,1,1,1,1,1,1,0,0,0,0,0,0},
       {1,0,0,0,1,1,1,1,1,1,1,0,0,0,0,0,0},
       {1,0,0,0,0,1,1,1,1,1,1,1,1,1,1,1,1},
       {0,1,1,0,0,1,1,1,0,1,1,1,1,1,1,1,1},
       {1,1,1,1,1,0,1,1,1,1,1,0,0,0,0,0,0},
       {1,1,1,1,1,1,0,1,1,1,0,1,1,1,1,1,1},
       {1,1,1,1,1,1,1,0,1,1,1,0,0,0,0,0,0},
       {0,1,1,1,0,1,1,1,0,1,1,0,0,0,0,0,0},
       {1,1,1,1,1,1,1,1,1,0,1,0,0,0,0,0,0},
       {1,1,1,1,1,1,0,1,1,1,0,1,1,1,1,1,1},
       {0,0,0,1,1,0,1,0,0,0,1,0,2,2,0,3,0},
       {0,0,0,1,1,0,1,0,0,0,1,2,0,0,3,0,2},
       {0,0,0,1,1,0,1,0,0,0,1,3,0,0,2,2,0},
       {0,0,0,1,1,0,1,0,0,0,1,0,2,3,0,0,3},
       {0,0,0,1,1,0,1,0,0,0,1,2,0,3,0,0,2},
       {0,0,0,1,1,0,1,0,0,0,1,0,3,0,2,3,0}}
      {
          \foreach [count=\c] \cell in \row
          {
              \ifnum\cell=1
                  \draw[arc1] (p\c) edge (p\r);
              \fi
              \ifnum\cell=2
                  \draw[arc] (p\c) edge (p\r);
              \fi
              \ifnum\cell=3
                  \draw[arc,dashed,red] (p\c) edge (p\r);
              \fi
          }
      }
      \end{pgfonlayer}
      \end{tikzpicture}
      }
}
\llap{\parbox[b]{2.9in}{(a)\\ \rule{0ex}{1.3in}}}
\hfil
\subfloat[]{
  \begin{tikzpicture}[vertex/.style={draw,circle,thin}]
      \node[anchor=south west,inner sep=0] (image) at (0,0,0) {\includegraphics[width=.25\textwidth]{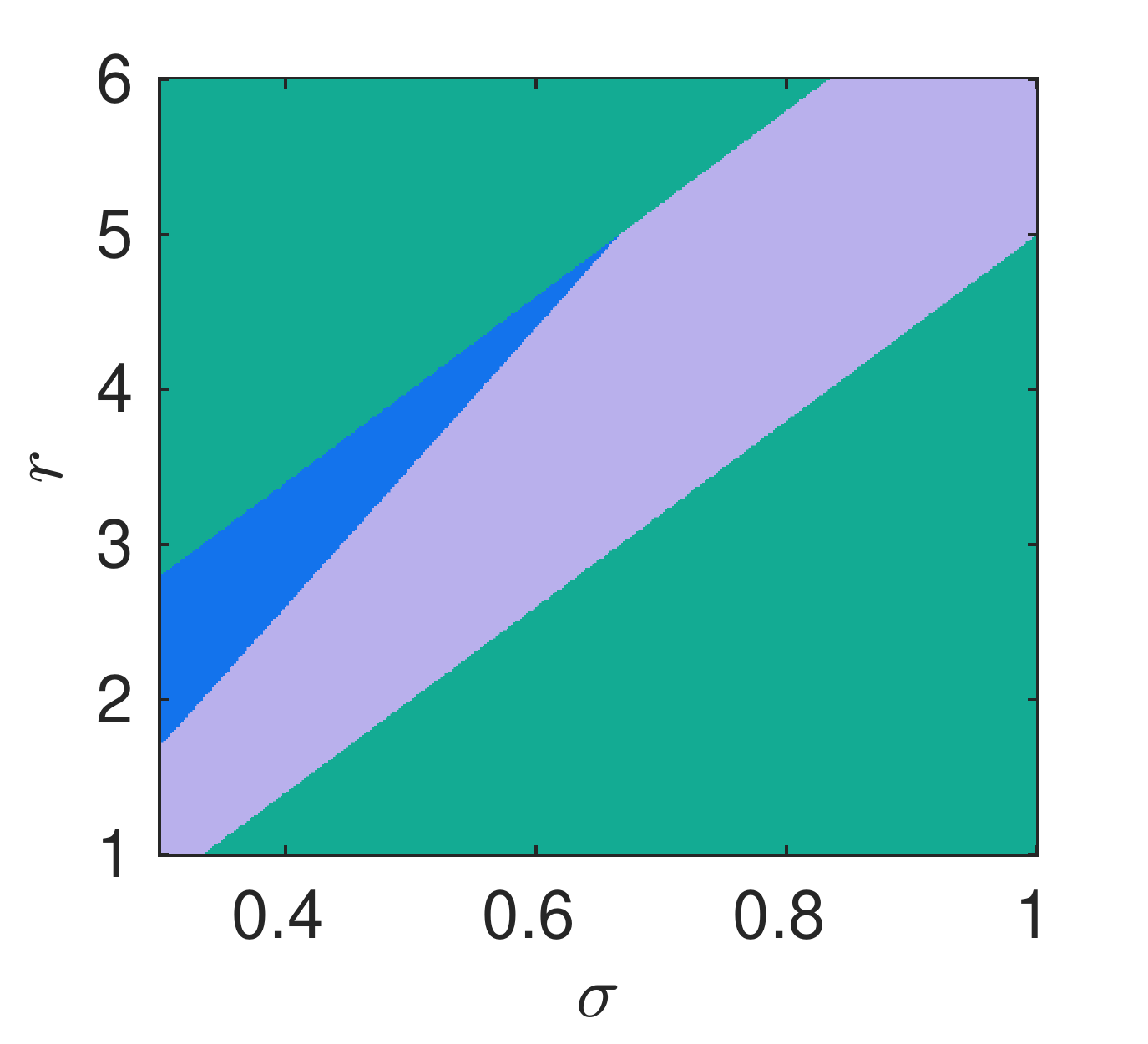}};
      \begin{scope}[x={(image.south east)},y={(image.north west)}]
        %\draw[dashed,thick] (.14,.78) edge (.91,.78);
        \node[] at (.55,.6) {AISync};
      \end{scope}
  \end{tikzpicture}
}
\llap{\parbox[b]{3.6in}{(b)\\ \rule{0ex}{1.3in}}}
\hfil
\subfloat[]{
  \begin{tikzpicture}[vertex/.style={draw,circle,thin}]
      \node[anchor=south west,inner sep=0] (image) at (0,0,0) {\includegraphics[width=.25\textwidth]{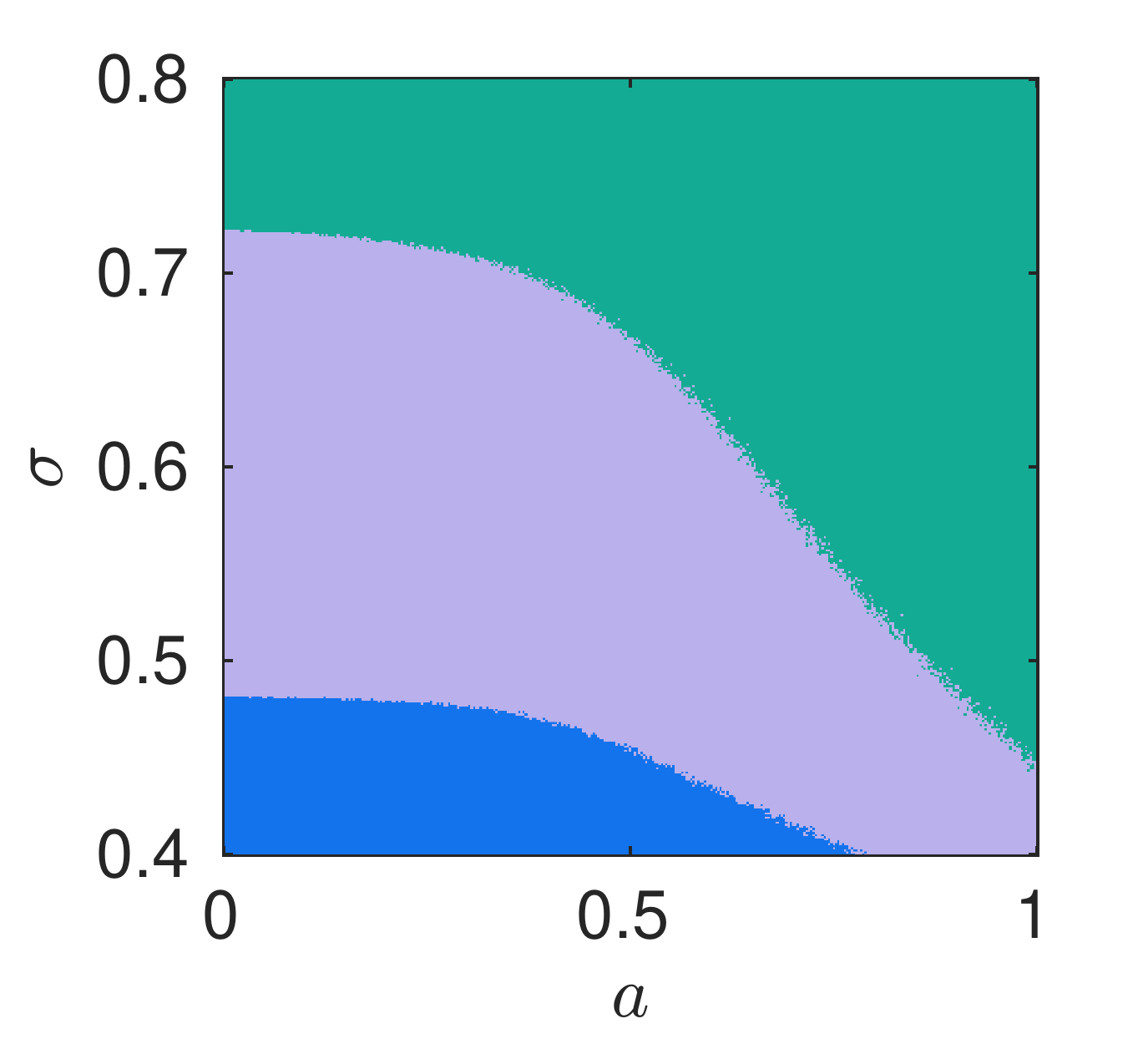}};
      \begin{scope}[x={(image.south east)},y={(image.north west)}]
        %\draw[dashed,thick] (.2,.37) edge (.9,.37);
        \node[] at (.45,.5) {AISync};
      \end{scope}
  \end{tikzpicture}
}
\llap{\parbox[b]{3.6in}{(c)\\ \rule{0ex}{1.3in}}}
\vspace{-6mm}
\caption{Cluster synchronization stabilized by breaking the structural symmetry of the cluster.
(a) Random network with five symmetry clusters (grouped nodes), in which the cluster considered in our examples is highlighted (magenta nodes).
Upon removal of the marked links (red), this cluster becomes optimally synchronizable but non-symmetric.
(b, c) Stability diagram in the  $r\times \sigma$ space for Bernoulli map oscillators (b) and in the $\sigma\times a$ space for H\'enon map oscillators (c).
The different colors mark regions in which synchronization is stable in both clusters (blue), unstable for both clusters (green), and unstable for the symmetric 
cluster but stable for the asymmetric one (purple), as determined by our calculation of the MTLE.
}
\label{fig:s1}
\end{figure}

To further demonstrate that the phenomenon we describe is common across different nodal dynamics 
and network structure, we present two additional examples.
For both examples we consider a random network with five symmetry clusters, as shown in \cref{fig:s1}(a). 
Within this network, we focus on the highlighted symmetry cluster (magenta nodes), which in isolation corresponds to the second symmetry cluster in \cref{tbl:tbl1}, and we contrast its synchronizability with that of the non-symmetric cluster generated by removing a subset of its links (red links).
 
We first consider this system when the nodes are equipped with dynamics of a Bernoulli map,
\begin{equation}
    x_i(t+1) = r\,x_i(t) - \sigma \sum_{j=1}^n L_{ij} x_j(t) \;\; \text{mod} \;\; 2\pi,
\end{equation}
which, for being piecewise linear and one dimensional, is  arguably one of the simplest possible nodal dynamics that one can consider in an oscillator network.
Despite its  simplicity, this system exhibits a rich stability diagram in the  $r \times \sigma$ parameter space, including a wide region in which synchronization is stable for the non-symmetric cluster but unstable for the symmetric one, as shown in \cref{fig:s1}(b). 
For $r \ge 5$, in particular, synchronization in the symmetric cluster is unstable for {\it any} coupling strength $\sigma$. 
Topological control is particularly valuable in this case as it allows for stability that would be impossible by merely adjusting $\sigma$ in the original cluster. 

As an illustration of higher dimensional nonlinear nodal dynamics, we also consider the system in \cref{fig:s1}(a) when equipped with the dynamics of a H\'enon map,
\begin{equation}
  \begin{cases}
    x_i(t+1) = 1 - a\,x_i^2(t) + y_i(t) - \sigma \sum_{j=1}^n L_{ij} y_j(t), \\
    y_i(t+1) = b\,x_i(t),
  \end{cases}
\end{equation}
where the variables $x_i$ and $y_i$ are defined on a torus and limited to $[-2,2]$; the coupling between oscillators are through the $y_i$ variables.
As shown in \cref{fig:s1}(c), fixing $b=0.3$ and calculating the stability diagram in the $\sigma \times a$ parameter space, once again we identify a wide region in which the non-symmetric cluster exhibits stable synchronization whereas the symmetric one does not. 

As illustrated by these and other systems we have studied in detail, in general a significant portion of the parameter space is occupied by a region in which synchronization is not stable for the symmetric cluster but it becomes stable when the structure of the cluster is optimized, which in turn goes in tandem with breaking its symmetry under the given constraints.
These examples also further illustrate the excellent agreement between direct simulations and theoretical predictions observed throughout.

\newpage

\section{Robustness of structural \texorpdfstring{AIS\MakeLowercase{ync}}{AISync}}

\begin{figure}[h]
\centering
\subfloat[]{
  \begin{tikzpicture}[vertex/.style={draw,circle,thin}]
      \node[anchor=south west,inner sep=0] (image) at (0,0,0) {\includegraphics[width=.45\textwidth]{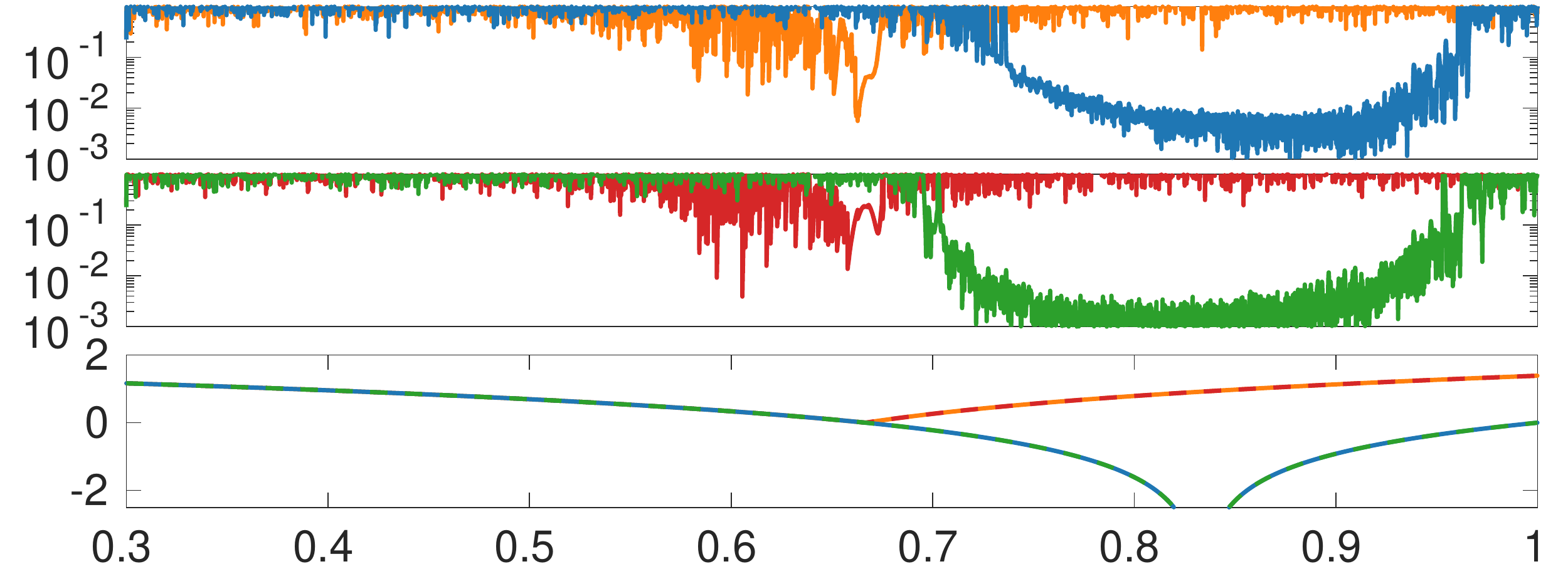}};
      \begin{scope}[x={(image.south east)},y={(image.north west)}]
        \node (rect) at (.53,.18) [fill=gray,minimum width=73mm,minimum height=4mm,fill opacity=0.2] {};
        \node () at (.55,-.05) {$\sigma$};
        \node[rotate=90] () at (-.01,.25) {\scriptsize MTLE};
        \node[rotate = 90] () at (-.02,.8) {$\Delta$};
        \node[rotate = 90] () at (-.02,.6) {$\Delta$};
        \draw[BurntOrange,thick] (.13,.9) edge (.18,.9);
        \node[] () at (.3,.9) {symmetric};
        \draw[MidnightBlue,thick] (.13,.8) edge (.18,.8);
        \node[] () at (.3,.8) {optimized};
        \draw[Red,thick] (.13,.6) edge (.18,.6);
        \node[] () at (.3,.6) {symmetric};
        \draw[ForestGreen,thick] (.13,.5) edge (.18,.5);
        \node[] () at (.3,.5) {optimized};
        \draw[dashed,thick] (.78,.1) edge (.78,1);
      \end{scope}
  \end{tikzpicture}
}
\llap{\parbox[b]{6.6in}{(a)\\ \rule{0ex}{1.2in}}}
\subfloat[]{
  \begin{tikzpicture}[vertex/.style={draw,circle,thin}]
      \node[anchor=south west,inner sep=0] (image) at (0,0,0) {\includegraphics[width=.4\textwidth]{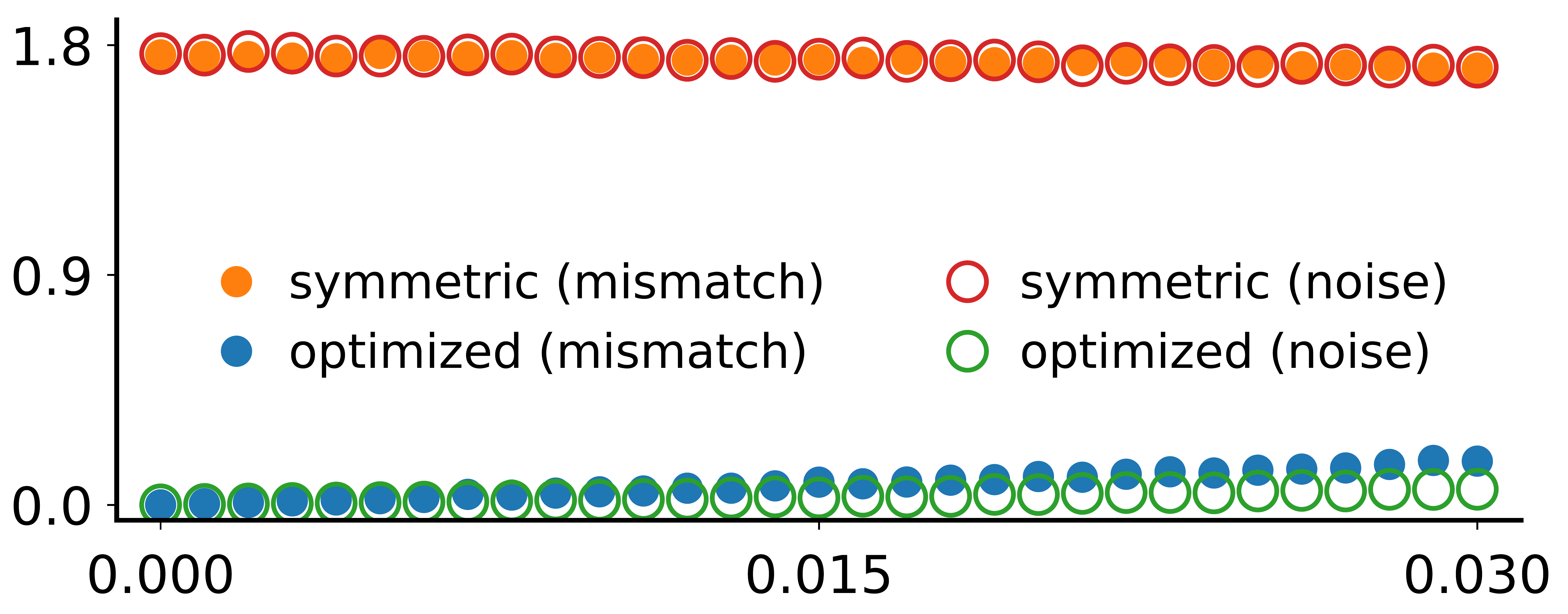}};
      \begin{scope}[x={(image.south east)},y={(image.north west)}]
        \node () at (.55,-.03) {$\xi$};
        \node[rotate = 90] () at (-.02,.55) {$\langle\Delta\rangle$};
      \end{scope}
  \end{tikzpicture}
}
\\[-1.5em]
\subfloat[]{
  \begin{tikzpicture}[vertex/.style={draw,circle,thin}]
      \node[anchor=south west,inner sep=0] (image) at (0,0,0) {\includegraphics[width=.45\textwidth]{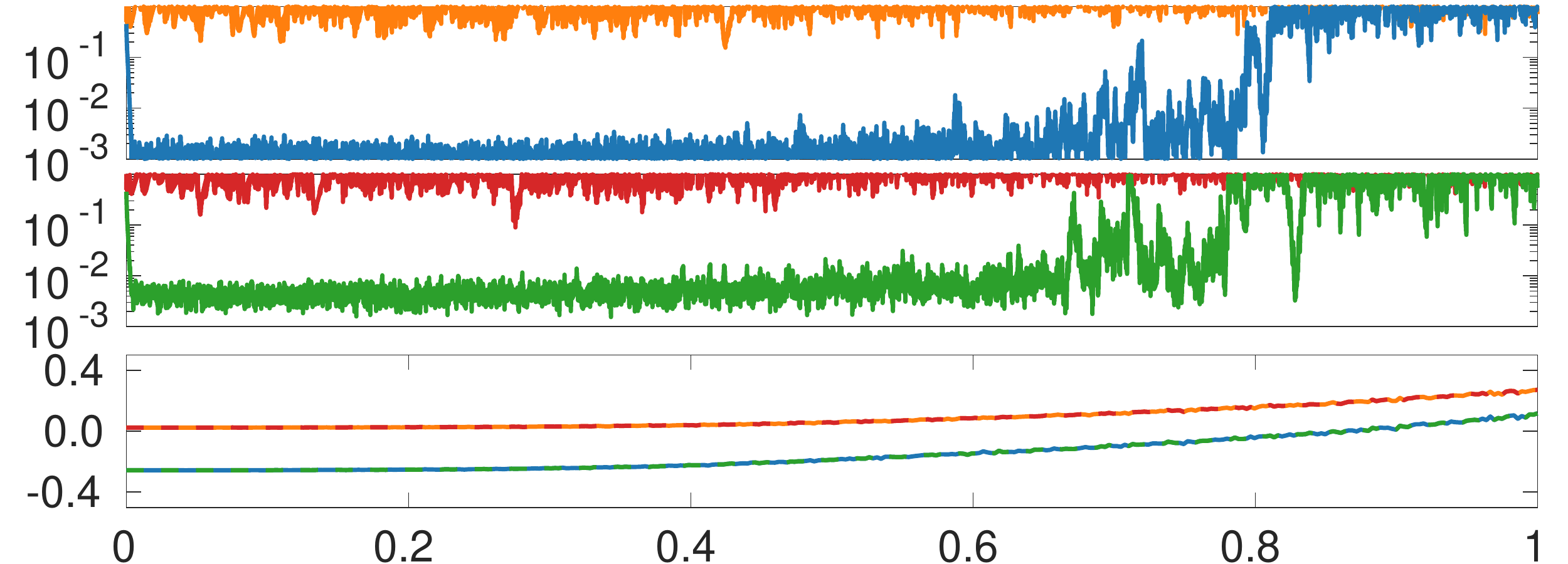}};
      \begin{scope}[x={(image.south east)},y={(image.north west)}]
        \node (rect) at (.53,.17) [fill=gray,minimum width=73mm,minimum height=4mm,fill opacity=0.2] {};
        \node () at (.55,-.05) {$a$};
        \node[rotate=90] () at (-.01,.25) {\scriptsize MTLE};
        \node[rotate = 90] () at (-.02,.8) {$\Delta$};
        \node[rotate = 90] () at (-.02,.6) {$\Delta$};
        \draw[dashed,thick] (.44,.1) edge (.44,1);
      \end{scope}
  \end{tikzpicture}
}
\llap{\parbox[b]{6.6in}{(b)\\ \rule{0ex}{1.2in}}}
\subfloat[]{
  \begin{tikzpicture}[vertex/.style={draw,circle,thin}]
      \node[anchor=south west,inner sep=0] (image) at (0,0,0) {\includegraphics[width=.4\textwidth]{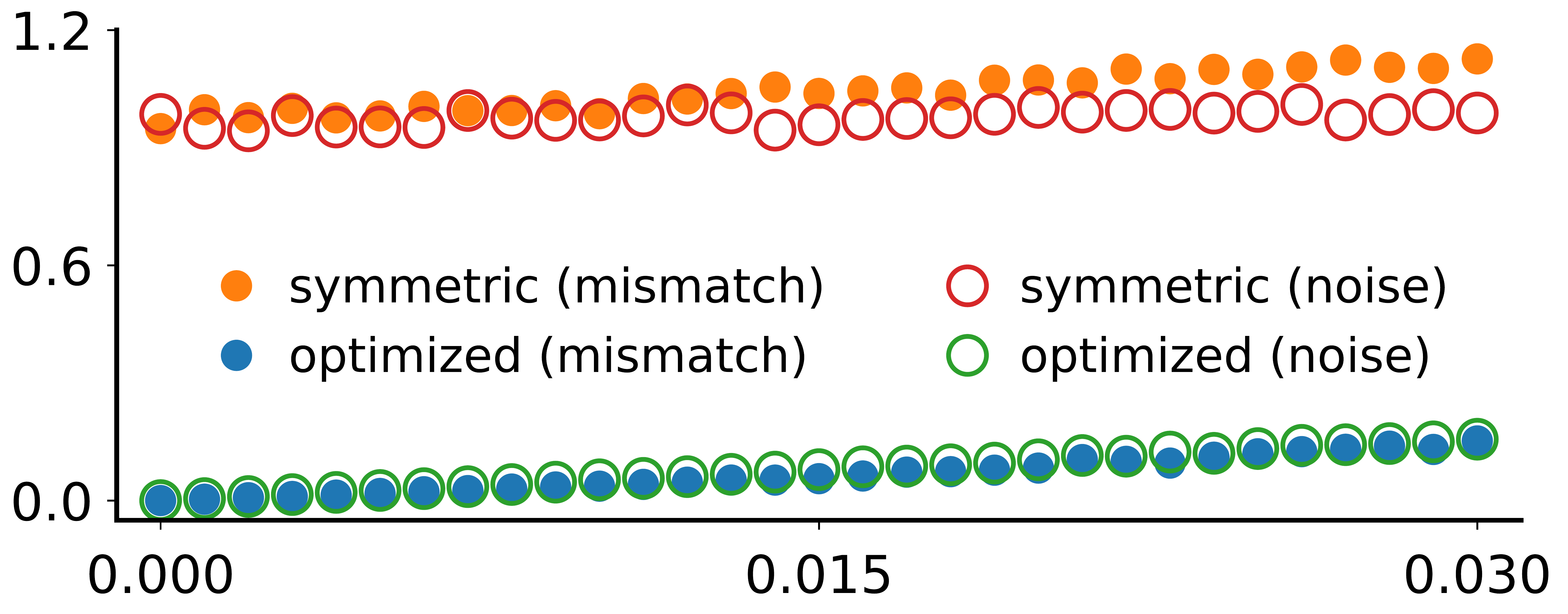}};
      \begin{scope}[x={(image.south east)},y={(image.north west)}]
        \node () at (.55,-.03) {$\xi$};
        \node[rotate = 90] () at (-.02,.55) {$\langle\Delta\rangle$};
      \end{scope}
  \end{tikzpicture}
}
\\[-1.5em]
\subfloat[]{
  \begin{tikzpicture}[vertex/.style={draw,circle,thin}]
      \node[anchor=south west,inner sep=0] (image) at (0,0,0) {\includegraphics[width=.45\textwidth]{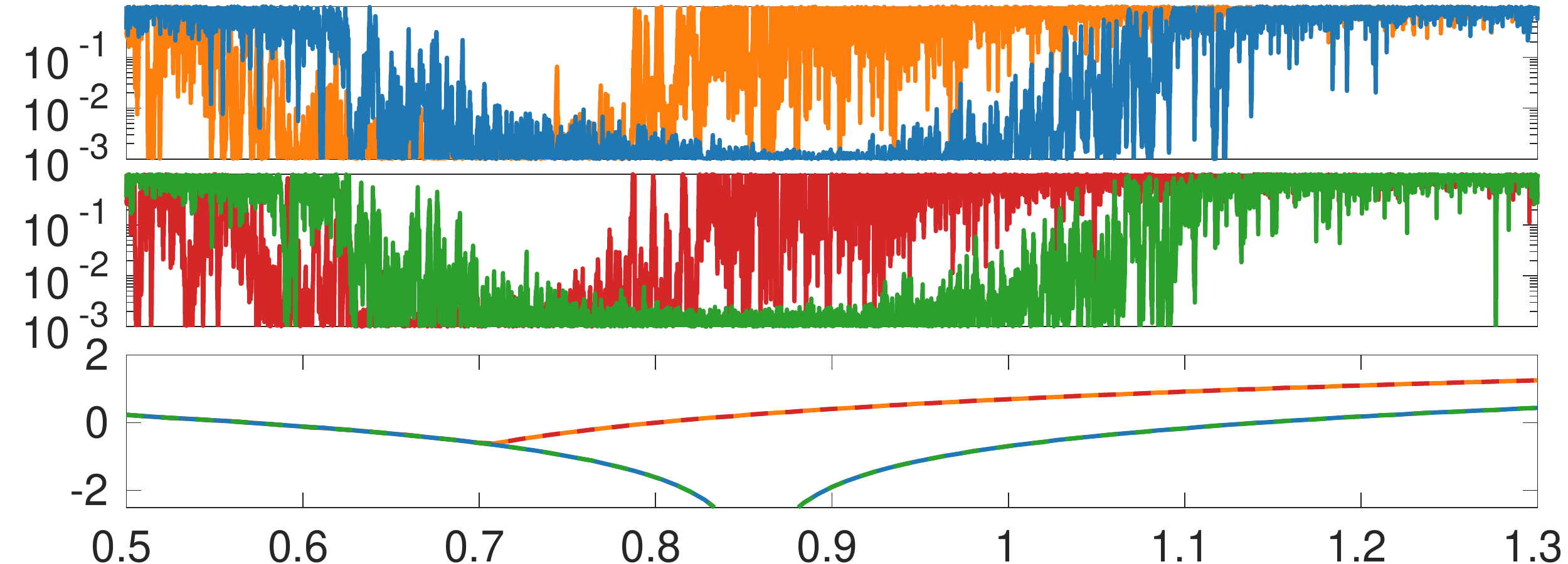}};
      \begin{scope}[x={(image.south east)},y={(image.north west)}]
        \node (rect) at (.53,.18) [fill=gray,minimum width=73mm,minimum height=4mm,fill opacity=0.2] {};
        \node () at (.55,-.05) {$\sigma$};
        \node[rotate=90] () at (-.01,.25) {\scriptsize MTLE};
        \node[rotate = 90] () at (-.02,.8) {$\Delta$};
        \node[rotate = 90] () at (-.02,.6) {$\Delta$};
        \draw[dashed,thick] (.53,.1) edge (.53,1);
      \end{scope}
  \end{tikzpicture}
}
\llap{\parbox[b]{6.6in}{(c)\\ \rule{0ex}{1.2in}}}
\subfloat[]{
  \begin{tikzpicture}[vertex/.style={draw,circle,thin}]
      \node[anchor=south west,inner sep=0] (image) at (0,0,0) {\includegraphics[width=.4\textwidth]{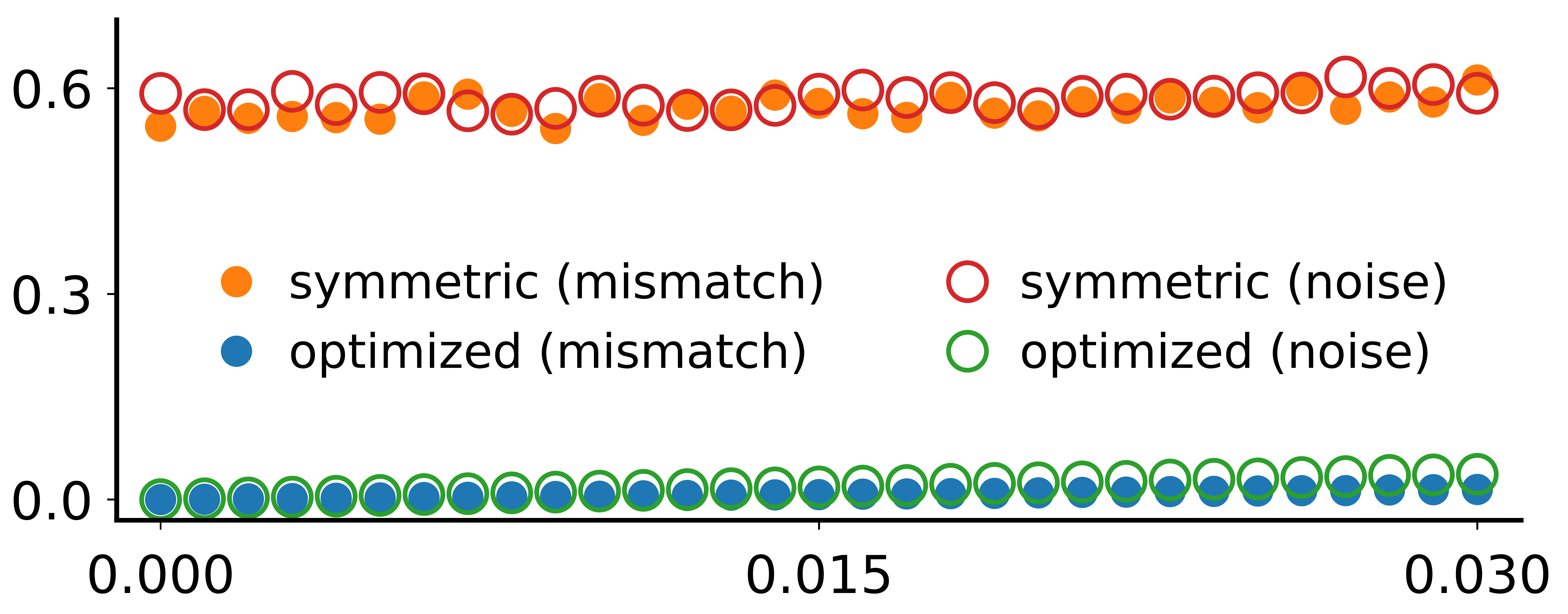}};
      \begin{scope}[x={(image.south east)},y={(image.north west)}]
        \node () at (.55,-.03) {$\xi$};
        \node[rotate = 90] () at (-.02,.55) {$\langle\Delta\rangle$};
      \end{scope}
  \end{tikzpicture}
}
\vspace{-6mm}
\caption{
Robustness of structural AISync against noise and oscillator heterogeneity, demonstrated for (a) the network of Bernoulli maps in \cref{fig:s1}, (b) the network of H\'enon maps in \cref{fig:s1}, and (c) the network of optoelectronic oscillators in \cref{fig:4}. 
Upper left panels: direct simulations with random parameter mismatch at magnitude $\xi=10^{-3}$. 
Middle left panels: direct simulations with noise at intensity $\xi=10^{-3}$.
Lower left panels: MTLE of the synchronized state in the symmetry cluster and optimized cluster. 
Right panels: dependence of the average synchronization error $\langle\Delta\rangle$ on $\xi$, when the system parameters are fixed at the value indicated by the dashed lines on the left.
%Three systems are analyzed:
%(a) network of Bernoulli maps from \cref{fig:s1};
%(b) network of H\'enon maps from \cref{fig:s1};
%(c) network of optoelectronic oscillators described in \cref{fig:4}.
}
\label{fig:s4}
\end{figure}

In order to demonstrate the robustness of structural AISync, we perform direct simulations of three different systems in the presence of Gaussian noise or random oscillator heterogeneity; the results are summarized in \cref{fig:s4}.
The three systems include the Bernoulli maps and H\'enon maps studied in Sec.~S4, as well as the optoelectronic oscillators from the main text.

In \cref{fig:s4}(a), we fix the parameter of the Bernoulli map to be $r=5$ and slowly increase the coupling strength $\sigma$ from 0.3 to 1.
For the trajectories in the upper left panel, a random mismatch of magnitude $\xi = 10^{-3}$ is introduced to the oscillator parameter $r$; 
For the trajectories in the middle left panel, the oscillators are subject to Gaussian noise with zero mean and standard deviation equal to $\xi = 10^{-3}$ (approximately the noise intensity in the experimental system).
Despite the noise and oscillator heterogeneity, the synchronization error $\Delta$ match well with the prediction based on the MTLE calculations shown in the lower left panel.
We investigate the dependence of the time-averaged synchronization error $\langle\Delta\rangle$ on the magnitude $\xi$ of noise/mismatch in the right panel, where $\sigma$ is fixed at 0.85 (corresponding to the dashed line on the left).

The same analysis is performed for the H\'enon maps in \cref{fig:s4}(b) and for the optoelectronic oscillators in \cref{fig:s4}(c).
For the H\'enon maps, mismatch is introduced in the parameter $b$, whose homogeneous value is set to $b=0.3$, for coupling strength fixed at $\sigma=0.5$.
For the optoelectronic oscillators, mismatch is introduced in the parameter $\beta$, whose homogeneous value is set to $\beta=6$.
It can be seen that in all three cases structural AISync is robust against both noise and oscillator heterogeneity.

\end{document}